\documentclass{aa}  

\usepackage{graphicx}
\usepackage{txfonts}

\usepackage{color}
\usepackage{ulem}

\usepackage{natbib}

\bibpunct{(}{)}{;}{a}{}{,} 

\begin{document}

   \title{On the consistent treatment of the quasi-hydrostatic layers in hot star atmospheres}
   \titlerunning{Treatment of the quasi-hydrostatic layers in hot star atmospheres}

   \subtitle{}

   \author{A. Sander\inst{1}
          \and
          T. Shenar\inst{1}
          \and
          R. Hainich\inst{1}
          \and
          A. G\'{\i}menez-Garc\'{\i}a\inst{2}
          \and
          H. Todt\inst{1}
          \and
          W.-R. Hamann\inst{1}
          }

   \institute{Institut f\"ur Physik und Astronomie, Universit\"at Potsdam,
              Karl-Liebknecht-Str. 24/25, D-14476 Potsdam, Germany\\
              \email{ansander@astro.physik.uni-potsdam.de, 
                     shtomer@astro.physik.uni-potsdam.de}
              \and
              Departamento de F\'{\i}sica, Ingenier\'{\i}a de Sistemas y Teor\'{\i}a de la Se\~{n}al,
              Universidad de Alicante, Apdo. 99, 03080 Alicante, Spain
              }

        \date{Received <date> / Accepted <date>}

\abstract{
Spectroscopic analysis remains the most common method to derive masses of massive stars, 
the most fundamental stellar parameter. While binary orbits and stellar pulsations can 
provide much sharper constraints on the stellar mass, these methods are only rarely applicable to massive stars. 
Unfortunately, spectroscopic masses of massive stars heavily depend on the 
detailed physics of model atmospheres.
}{
We demonstrate the impact of a consistent treatment of the radiative pressure on inferred gravities and 
spectroscopic masses of massive stars. Specifically, we investigate the contribution of line and continuum 
transitions to the photospheric radiative pressure.
We further explore the effect of model parameters, e.g.,\ abundances, on the deduced spectroscopic mass. Lastly, we compare our results with the 
plane-parallel TLUSTY code, commonly used for the analysis of massive stars with 
photospheric spectra.
}{
  We calculate a small set of O-star models with the Potsdam Wolf-Rayet (PoWR)
  code using different approaches for the quasi-hydrostatic part. These models 
  allow us to quantify the effect of accounting for the radiative pressure consistently. We further use 
  PoWR models to show how the Doppler widths of line profiles and abundances of elements such as iron affect 
  the radiative pressure, and, as a consequence, the derived spectroscopic masses.
}{
  Our study implies that errors  on the order of a factor of two in 
  the inferred spectroscopic mass are to be expected when neglecting the contribution of line and continuum transitions
  to the radiative acceleration in the photosphere. 
  Usage of implausible microturbulent velocities, or the neglect of 
  important opacity sources such as Fe, may result in errors of approximately 50\%\ in the spectroscopic mass.
  A comparison with TLUSTY model atmospheres reveals a very good agreement with PoWR at the limit of low mass-loss rates.
}
{}


\maketitle


\section{Introduction}
  \label{sec:intro}
  
  The initial mass of a star determines its evolutionary
  path, and is thus considered  one of the most fundamental 
  stellar parameters. Yet stellar masses derived for massive
  stars via spectral analyses
  are generally prone to large uncertainties, 
  greatly hampering an accurate calibration of stellar masses 
  with their spectral types and evolutionary status. 
  The so-called ``mass discrepancy'' problem, which arises when comparing stellar masses 
  obtained from spectroscopy, to orbital, wind, and evolutionary models, has been of concern to stellar physicists for 
  a few decades \citep{Herrero+1992, Repolust+2004, Massey+2012}. 
  While recent studies suggest a solution of this problem over the years 
  \citep{Weidner+2010, Markova+2014}, discrepancies still exist, especially 
  in the range of giants to supergiants.
  
  In principle, orbital masses are independent of stellar atmosphere models and are therefore considered 
  to be more robust \citep[e.g.,][]{Torres+2011}. However,
  orbital masses are only attainable in  the rare case of binary systems with well-constrained inclination, usually 
  owing to eclipses. Immense progress has also been made in the field of asteroseismology, which allows the measurement of  stellar masses with very high accuracies from observed stellar pulsations.
  Unfortunately, the high variability in the outer layers of massive stars make the study of their pulsational behavior very difficult, 
  often hindering an effective implementation of asteroseismological methods to massive stars \citep[see recent review by][and references therein]{Aerts2014}. Indeed, spectroscopy remains the primary method to infer stellar masses for
  the majority of the massive stars.
  
  The spectroscopic mass of a star is derived from its radius $R_*$ 
  and surface gravity $g_\ast$ via $M_\ast = G^{-1}\,g_\ast\,R_*^{2}$, and therefore any uncertainties in 
$R_*$ and $g_\ast$ propagate into uncertainties in $M_\ast$. Except for the rare case where the 
  angular diameter of a star can be directly measured, 
  $R_*$ is derived from the luminosity 
  $L$ and effective temperature $T_\text{eff}$ of the star via the Stefan-Boltzmann 
  relation $R_\ast= \left(4\pi\sigma_\text{SB}\right)^{-1/2}\,L^{1/2}\,T_\text{eff}^{-2}$. 
  Since the effective temperature can in principle 
  be constrained with  decent accuracy,
  the main cause for uncertainty in the stellar radius is the error in the distance $d$, 
  which propagates in the error in $L$, according to $R_\ast \propto \sqrt{L} \propto d$.
  If the distance is well
  constrained and thus the spectroscopic radius is known, the stellar mass $M_\ast$
  is directly proportional to $g_\ast$ and thus all uncertainties in the gravity determination
  propagate directly into mass uncertainties.

  The gravity $g_*$ is by no means a directly measurable quantity.
  It is derived by comparing synthetic spectra from model atmospheres with observations. 
  The surface gravity 
  of a star determines the stratification of its atmospheric pressure, and most prominently affects 
  the pressure-broadened wings of hydrogen and helium lines. The determination of $g_*$ 
  thus relies on both the pressure broadening theory adopted, as well as the model atmosphere. 

  Different broadening theories can 
  lead to systematic differences of $\sim 0.15\,$dex 
  in inferred $\log g_\ast$ values, which alone corresponds to $\sim 40\%$ error 
  in the inferred spectroscopic mass. It is an even harder task, however, to constrain 
  the systematic errors that arise because of different assumptions and techniques 
  in stellar atmosphere codes. The radiative pressure in massive stars 
  depends on the stellar parameters
  as well as the opacity, which in turn depends on the elements, abundances, atomic data, and line Doppler widths.  
  As we illustrate in this study,
  $g_\ast$ is highly model-dependent, and systematic errors can easily occur if the radiative pressure 
  is not fully and consistently accounted for. 
  
   Two domains can be distinguished in the atmosphere of a massive
   star: a hydrostatic domain,
   where gravity is balanced by pressure (e.g., gas pressure,
   radiation pressure), and a wind domain, where the outward
   pressure exceeds gravity and the matter is accelerated.
   
  The spectra of O- and B-type stars are mostly formed in the outer
  layers of their quasi-hydrostatic domains. For the modeling of such stars, a detailed
  treatment of the hydrostatic regime is imperative. Specifically, 
  the contribution of line and continuum transitions to the 
  total radiative pressure in the quasi-hydrostatic domain is far from negligible. 
  Proper knowledge of the velocity field in the layers close to
  the stellar surface is a key ingredient for a better understanding of a variety
  of theoretical and observational phenomena 
  \citep[see, e.g.,][]{H1981,CR1993,OP1999,Cantiello+2009,SHT2014}.
  
  In this paper, we thoroughly document the current calculation of the radiative pressure 
  in the Potsdam Wolf-Rayet (PoWR) code \citep[see, e.g.][]{GKH2002,HG2003} and illustrate 
  the importance of accounting for it consistently.
  While we focus on the PoWR code, the major concepts and equations 
  are representative of the majority of current state-of-the-art stellar atmosphere codes, as we
   illustrate in a brief comparison. We further discuss and quantify
  the large impact of a proper hydrostatic treatment on inferred stellar parameters, and particularly 
  on the stellar mass $M_\ast$. We demonstrate the sensitivity of the 
  radiative pressure to the Doppler width of spectral lines and to the iron abundance. 
  Lastly, we compare our results with the plane-parallel TLUSTY O-grid 
  models from \citet{LH2003}, which are widely used for the analysis of hot stars with negligible winds.

  The structure of this paper is as follows: In Sect.\,\ref{sec:code}, we briefly summarize the main
    assumptions of the PoWR code and thoroughly discuss the treatment
  of the quasi-hydrostatic domain. The outcome of our test calculations
  are shown and discussed in
    Sect.\,\ref{sec:results}. In Sect.\,\ref{sec:tlusty}, we compare our
    spectra with corresponding TLUSTY models, before drawing the general conclusions in
   Sect.\,\ref{sec:conclusions}.

\section{The PoWR code}
  \label{sec:code}

\subsection{The Basics}
\label{sec:overview}

  The Potsdam Wolf-Rayet models describe atmospheres of spherically symmetric stars with a
  stationary outflow\footnote{For Wolf-Rayet stars, model grids are available online
  at \texttt{http://www.astro.physik.uni-potsdam.de/PoWR/}}. To achieve a consistent solution, the equations
  of statistical equilibrium and radiative transfer are iteratively
  solved to yield the population numbers without the approximation of a local thermodynamic equilibrium (non-LTE). 
  The radiative transfer is solved in the comoving frame, which avoids simplifications such as the 
  Sobolev approximation.
  After an atmosphere model is converged, the synthetic spectrum is 
  calculated via a formal integration along emerging rays.
  Some description of the PoWR code can be found in \citet{GKH2002}
  and \citet{HG2004}.
  The temperature stratification is updated iteratively to ensure energy conservation 
  in the expanding atmosphere, as described in \citet{HG2003}. 
  Recently, the PoWR code 
  has been extended by the so-called thermal balance-method, which
  goes back to ideas of \citet{HS1963} and \citet{H1963} and is
  described in detail for stellar atmospheres by \citet{KPP1999} and \citet{K2001}. 
  This method provides better numerical stability in optically thin domains. 
  
  In the comoving frame calculations during the non-LTE iteration, we assume that the line profiles
  are Gaussians with a constant Doppler broadening velocity $\varv_\text{dop}$, which approximately
  accounts for the thermal and turbulent velocity. A constant broadening velocity
  is a well-established simplification in comoving frame methods. While PoWR uses directly a
  velocity $\varv_\text{dop}$ as input, the ``Comoving Frame General'' (CMFGEN) code requires three 
  input parameters $T_\text{dop}$, 
  $A_\text{dop}$, and $\varv_\text{T}$ , which are then combined to a constant velocity
  $\varv_\text{dop} = \sqrt{2 \frac{k_\text{B} T_\text{dop}}{m_\text{H} A_\text{dop}} + \varv_\text{T}^2}$
  \citep[see, e.g.,][CMFGEN description]{MSH2002}. In the FASTWIND\footnote{acronym for ``fast analysis
  of stellar atmospheres with winds''} code a velocity similar to PoWR has to be
  given and is referred to just as mircroturbulence $\varv_\text{turb}$ \citep{Puls+2005}.
  
  The value of $\varv_\text{dop}$ is chosen such that it approximately reflects the order 
  of the averaged thermal speed combined with the microturbulence, i.e.,
  \begin{equation}
    \varv_\text{dop} \approx \sqrt{ \bar{\varv}_\text{th}^2 + \varv_\text{turb}^2 }
  .\end{equation}
  For O and B stars typical values for $\varv_\text{dop}$ range 
  between $10$ and $30\,$km/s. Only for stars without photospheric lines, such as classical Wolf-Rayet
  stars, higher values can be chosen  to speed up the calculations without 
  changing the emergent spectrum. The influence of $\varv_\text{dop}$ on an O-star model
  spectrum is discussed and illustrated in Sect.\,\ref{sec:resotherparam}.
   
  Pressure broadening is neglected during the comoving frame calculations, which is sufficient 
  for the current applications of the PoWR models. For stars
  with considerably higher values of $\log g_\ast$, e.g. , subdwarfs, specific codes, 
  such as the T{\"u}bingen Model-Atmosphere Package (TMAP) \citep[e.g.,][]{Werner+2003} exist which 
  include this effect. Some codes, such as the TLUSTY code, have the option to switch on pressure 
  broadening in the iteration if needed. In the formal integration in PoWR, detailed thermal, microturbulent 
  and pressure broadening are accounted for in a depth-dependent manner. 
  
  The basic parameters of a PoWR model are the stellar temperature $T_\ast$, luminosity $L$, mass-loss rate $\dot{M}$,
  surface gravity $g_*$ and the chemical abundances. The stellar temperature $T_\ast$ is defined as the effective temperature
  of a star with the luminosity $L$ and radius $R_\ast$ (referred to as the ``stellar radius''),  
  defined at the Rosseland continuum optical depth $\tau_\text{max} = 20$. The total Rosseland 
  optical depth $\tau_\text{Ross}(R_\ast)$ including lines is larger.
  To ensure $\tau_\text{max} = 20$ at the inner boundary, the velocity $\varv_\text{min} = \varv(R_\ast)$
  is iteratively adjusted. The surface gravity $g_\ast$ is 
  defined at the stellar radius $R_*$ via $g_\ast = G\,M_\ast\,R_\ast^{-2}$. For OB stars, the difference
  between the radius where $\tau_\text{max} = 20$ and the ``photospheric radius''  
  at $\tau_\text{Ross} = 2/3$ is usually very small. However,
  in supergiants the effective temperature can differ up to $\sim 1$\,kK between 
  these two points, and this difference in definition 
  is apparent when compared with other studies.
  
  The density stratification in the quasi-hydrostatic domain
  follows from an integration of the hydrostatic equation, 
  thoroughly discussed in Sect.\,\ref{sec:physics}.
  In the wind domain, the radial wind velocity $\varv(r)$ is usually prescribed 
  in the model by a so-called $\beta$-law 
  \begin{equation}
    \varv(r) = \varv_\infty \left( 1 - \frac{R_{\ast}}{r}
    \right)^\beta ,
   \label{eq:beta}
  \end{equation}  
  where $\varv_\infty$ is the terminal velocity of the wind, and $\beta$ 
  is a free input parameter whose value typically ranges between 
  $\beta = 0.6$ and $\beta = 2.0$ ,\citep[e.g.,][]{PVN2008}. With 
  the mass-loss rate $\dot{M}$ specified, the density stratification $\rho(r)$ in 
  the wind follows from the continuity equation
   \begin{equation}
    \label{eq:cont}
    \dot{M} = 4 \pi r^2 \varv(r)\,\rho(r) \text{.}
  \end{equation}
 
  In the calculation, we use complex model atoms, with a superlevel
  approach for iron group elements \citep[see][for details]{GKH2002}, 
  and an explicit set of quantum levels for all other elements. The detailed 
  chemical composition for our calculations is given in Sect.\,\ref{sec:results},
  together with the rest of the model parameters. 
   
  We do not use any clumping in the models and instead assume a
  smooth wind. The potential existence of clumping in the
  subsonic photosphere is a constant debate in the massive star community
  \citep[see, e.g.,][]{RO2002,OHF2007,Cantiello+2009,SO2013}.
  While clumping would significantly affect the derived absolute stellar
  parameters in
  any case, the goal of this work is to demonstrate effects that are 
  independent of any clumping formalism. Therefore, we refrain from assuming
  a particular clumping approximation and calculate our test models 
  (see Sect.\,\ref{sec:results}) with a smooth wind.

\subsection{The quasi-hydrostatic domain}
  \label{sec:physics}
  
  The Potsdam Wolf-Rayet (PoWR) model atmosphere code was originally 
   developed for WR stars where the emergent
  spectrum is formed almost exclusively in the stellar wind. For these
  objects, it is sufficient to treat the quasi-hydrostatic domain of WR 
  stars with a simple barometric formula using a constant scale height
  \begin{equation}
    \label{eq:hconst}
    H_\text{c} := \frac{\frac{\mathcal{R} T_\ast}{\mu} + \varv_\text{turb}^2}{g_\text{eff}~R_\ast},
  \end{equation}  
  in units of $R_\ast$ while ensuring a smooth transition of the velocity field and its 
  gradient between the two quasi-hydrostatic domain and the wind. Here, $g_\text{eff}$ is the 
   gravity corrected for radiative pressure in the hydrostatic domain 
    (see details below), $\mu$ the mean particle mass (including electrons)
  in units of the hydrogen atom mass $m_\text{H}$, and $\varv_\text{turb}$ the turbulent velocity, 
  which is a free input parameter. The parameter $\mathcal{R}$ denotes 
  the specific gas constant for hydrogen, i.e., $\mathcal{R} = k_\text{B} / m_\text{H}$.
  
  For a proper treatment of O- and B-star atmospheres, the barometric formula with a
  constant scale height is not an appropriate solution of the hydrostatic
  equation. The hydrostatic equation is one of the fundamental equations of
  stellar structure. For massive stars,
  one has to account for the outward radiative force acting against 
  gravitation, yielding the radial stratification of
  pressure $P(r)$ in the hydrostatic domain,
  \begin{equation}
    \label{eq:hydrostat}
    \frac{\mathrm{d}P}{\mathrm{d}r} = - \rho(r) g(r) \left[ 1 - \Gamma(r) \right].
  \end{equation}  
  Here, $\rho(r)$ is the mass density, $g(r) = G\,M_* / r^2$ the
  gravity, and $\Gamma(r)$ the ratio between the outward radiative acceleration and gravitational acceleration.  
  The term $\left[1 - \Gamma(r)\right]$ describes the
  effective reduction of the gravity due radiative
  pressure, as discussed below.
  
 With the assumption of an ideal gas, the pressure $P$ can be expressed by
  \begin{align}
    P(r) & = \rho(r) \left[ \frac{\mathcal{R} T(r)}{\mu(r)} + \varv^2_\text{turb} \right] \\
         \label{eq:idgas}
         & = \rho(r) \left[ a^2(r)  + \varv^2_\text{turb} \right] \text{,}
  \end{align}
  where $T$ is the electron temperature.
  In the second line we further introduce the sound speed
  \begin{equation}
    a(r) := \sqrt{\frac{\mathcal{R} T(r)}{\mu(r)}}
  \end{equation}
  in order to simplify the expression.
  The turbulent velocity $\varv_\text{turb}$ is a free depth-independent input 
  parameter reflecting a possible microturbulence. While in the formal integral the
  microturbulence is combined with the actual thermal velocity of each element to obtain the precise
  depth-dependent Doppler broadening
  velocity, it is not directly connected to the value of $\varv_\text{dop}$ used in the
  comoving-frame calculation. To avoid physically inconsistent situations, the
  value of $\varv_\text{dop}$ should be higher than for $\varv_\text{turb}$. So far this
  has been ensured by the user, but we are planning to implement a more detailed treatment
  of microtubulence in the comoving-frame calculations, also allowing for depth-dependent 
  changes. 
  
  The current implementation allows us to compare our OB-type
  models with those of other stellar atmosphere codes, as they adopt similar approaches (cf. 
  Table \ref{tab:codecmp}). Stellar atmosphere analyses with several codes 
  \citep[see, e.g.,][]{Massey+2013} have demonstrated that microturbulent velocities of the order of
  $10$ to $20\,$km/s help to reproduce the observed spectral lines in certain parameter
  regimes. It is an ongoing debate whether such velocities represent a real turbulent motion
  in the photosphere, as originally suggested for hot stars by \citet{SE1934} and prominently
  reintroduced by \citet{HHA1991}, or if they are rather a
  ``fudge factor''. Related to this open question is the discussion whether such a turbulent term 
  should be included in the hydrostatic equation or not. In the PoWR code the term is included,
  but because of its currently depth-independent implementation there are no additional derivatives
  occuring and it merely leads to an offset of the sound speed.

  For simplicity, we will set $\varv_\text{turb} = 0$\,km/s
  in the following calculations, but the full result can always be recovered by replacing
  $a^2$ with $a^2 + \varv^2_\text{turb}$.   
  Upon dividing Eq.\,(\ref{eq:hydrostat}) by $-\rho$, the term on the left 
  side, which we will refer to as 
  \begin{equation}
    \label{eq:apressdef}
     a_\text{press}(r) := -\frac{1}{\rho} \frac{\mathrm{d}P}{\mathrm{d}r}
  \end{equation} 
   describes the outward acceleration
  due to gas pressure and turbulent motion. Thus the hydrostatic equation can be
  written as
  \begin{equation}
    \label{eq:hystapress}
    a_\text{press}(r) = g(r) \left[ 1 - \Gamma(r) \right]\text{,}
  \end{equation}
  illustrating that in the hydrostatic domain the outward acceleration due to gas pressure has 
  to balance gravity reduced by the radiative acceleration. In a strictly hydrostatic domain,
  we have therefore no net velocity, i.e., no stellar wind. 
  
  For an accurate description of an expanding stellar atmosphere, i.e., with $\varv \neq 0$, one
  would have to use the \textit{\textup{hydrodynamic}} equation, which can be written for a stationary,
  symmetric outflow in the following form:
  \begin{align}
    \label{eq:hydrodyn}
      \frac{1}{\rho} \frac{\mathrm{d}P}{\mathrm{d}r} +  \varv \frac{\mathrm{d}\varv}{\mathrm{d}r} & =  - g(r) \left[ 1 - \Gamma(r) \right.].
  \end{align}
  This means there is only one additional term in Eq.\,(\ref{eq:hydrodyn}) compared to the 
  hydrostatic Eq.\,(\ref{eq:hydrostat}). This inertia term $a_\text{mech}(r) := \varv \frac{\mathrm{d}\varv}{\mathrm{d}r}$
  is of fundamental importance in the outer wind, but becomes negligible quickly below the sonic point.
  This also holds for a $\beta$-type velocity law with $\beta > 0.5$ where $a_\text{mech}$ approaches zero for $r \rightarrow R_\ast$.
  Thus in the subsonic domain the hydrodynamic equation (\ref{eq:hydrodyn}) transitions into
  the hydrostatic form (\ref{eq:hydrostat}) and we obtain a quasi-hydrostatic stratification.
  
  In such a quasi-hydrostatic situation, we can still have a nonvanishing velocity, even though
  its value is subsonic and, as we will see in the comparison with purely hydrostatic models in Sect.\,\ref{sec:tlusty},
   negligible for the observed spectrum. Furthermore this small velocity still fulfills the
  equation of continuity (\ref{eq:cont}). This can be used to obtain the consistent solution
  for the velocity field in the quasi-hydrostatic domain. First, we replace the pressure gradient 
  in $a_\text{press}$ with the  help of Eq.\,(\ref{eq:idgas}),
  \begin{align} 
     a_\text{press} & = -\frac{1}{\rho} \frac{\mathrm{d}P}{\mathrm{d}r} \\
                  & = - \frac{\mathrm{d}a^2}{\mathrm{d}r} - \frac{a^2}{\rho} \frac{\mathrm{d}\rho}{\mathrm{d}r}.
  \end{align} 

  We then eliminate the density in the second term by using the equation of continuity (\ref{eq:cont})
  and thus write
  \begin{align} 
    a_\text{press} & = - \frac{\mathrm{d}a^2}{\mathrm{d}r} - a^2 r^2 \varv \frac{\mathrm{d}}{\mathrm{d}r} \left(\frac{1}{r^2 \varv}\right) \\
      \label{eq:apresshydro}
      & = - \frac{\mathrm{d}a^2}{\mathrm{d}r} + \frac{2 a^2}{r} + \frac{a^2}{\varv} \frac{\mathrm{d}\varv}{\mathrm{d}r} \text{.}
  \end{align} 
   
  Finally, by combining Eqs.\,(\ref{eq:hystapress}) and (\ref{eq:apresshydro}), 
  we obtain an equation for the velocity gradient in the quasi-hydrostatic regime,
  \begin{equation}
    \label{eq:vgrad}
     \frac{\mathrm{d}\varv}{\mathrm{d}r} = \frac{\varv}{a^2} \left[ \frac{G M_\ast}{r^2} \left(1 - \Gamma(r)\right) -  \frac{2      a^2}{r} + \frac{\mathrm{d}a^2}{\mathrm{d}r} \right]
  .\end{equation}
  
  In principle, given a value
  $\varv(R_*) = \varv_\text{min}$ at the inner
  boundary, one can obtain the velocity field in the quasi-hydrostatic part
  via direct integration of this equation. In practice, we transform this
  equation and split off the main exponential trend as this turned out to
  work better in terms of numerical stability. Thus our solution for
  the quasi-hydrostatic velocity field is
  \begin{equation}
    \label{eq:vsol}
    \varv(r) = \varv_\text{min} \frac{a^2(r)}{a^2(R_\ast)} \frac{R_\ast^2}{r^2} \exp\left( \frac{r - R_\ast}{H_\text{c}} - b(r) \right)  
  ,\end{equation}
  with a yet to be determined function $b(r)$. We now calculate
  the derivative of Eq.\,(\ref{eq:vsol}) with respect to r,
  \begin{equation}
    \frac{\mathrm{d}\varv}{\mathrm{d}r} = - \frac{2\varv}{r} + 2\frac{\varv}{a^2}\frac{\mathrm{d}a^2}{\mathrm{d}r} + \varv \left[ \frac{1}{H_\text{c}} - \frac{\mathrm{d}b}{\mathrm{d}r} \right] \text{,}
  \end{equation}
  combine this with Eq.\,(\ref{eq:vgrad}), and obtain
  \begin{align}  
  - \frac{2}{r} + \frac{1}{a^2}\frac{\mathrm{d}a^2}{\mathrm{d}r} + \frac{1}{H_\text{c}} - \frac{\mathrm{d}b}{\mathrm{d}r}
   & = \frac{G M_\ast}{a^2 r^2} \left(1 - \Gamma(r)\right) -  \frac{2}{r} + \frac{1}{a^2} \frac{\mathrm{d}a^2}{\mathrm{d}r} .\\
    \label{eq:dbdreq}
   \frac{1}{H_\text{c}} - \frac{\mathrm{d}b}{\mathrm{d}r} & = \frac{G M_\ast}{a^2 r^2} \left(1 - \Gamma(r)\right)
  \end{align}
  
  The righthand side of Eq.\,(\ref{eq:dbdreq}) is now defined as $H(r)^{-1}$, as it is analogous to the definition of $H_\text{c}$ (Eq.\,\ref{eq:hconst}). Thus the numerical integration of the velocity gradient is replaced by the integration of
  \begin{equation}
    \label{eq:dbdr} 
    \frac{\mathrm{d}b}{\mathrm{d}r} = \frac{1}{H_\text{c}} - \frac{1}{H(r)}\text{.}
  \end{equation}
  The required boundary value $b(R_\ast) = 0$ follows from the inner boundary, where it is required in Eq.\,(\ref{eq:vsol}) that $\varv(R_\ast) = \varv_\text{min}$.

  So far, we did not discuss what enters $\Gamma(r)$ and thus $H(r)$. In general,
  the letter $\Gamma$ is used to describe a ratio between a radiative acceleration and
  gravity. The most common use is the so-called ``electron gamma'',  
  \begin{equation}
    \label{eq:Gammae}
    \Gamma_\text{e} = \frac{a_\text{thom}(r)}{g(r)} = \frac{\sigma_\text{e}}{4\pi c m_\text{H} G} q_\text{ion}(r) \frac{L}{M_\ast},
  \end{equation}
  which accounts only for the radiative pressure because of scattering of free 
  electrons. Note that $r^2$ cancels out in the last expression, since the
  acceleration because of Thomson scattering is defined as 
  \begin{equation}
    \label{eq:athom}
    a_\text{thom}(r) := \frac{\sigma_\text{e} L}{4\pi c m_\text{H} r^2} q_\text{ion}(r),
  \end{equation}
  with the ionization parameter, 
  \begin{equation}
    q_\text{ion}(r) = m_\text{H} \frac{n_\text{e}(r)}{\rho(r)} = \frac{n_\text{e}(r)}{n_\text{tot}(r) \mathcal{A}}
  .\end{equation}
  The mean atomic mass $\mathcal{A}$ is constant in an atmosphere model. The two density factors can be replaced by the
  slowly varying mean particle mass $\mu(r) = \mathcal{A} \left( 1 + \frac{n_\text{e}(r)}{n_\text{tot}(r)} \right)^{-1}$
  such that
  \begin{equation}
    q_\text{ion}(r) = \frac{1}{\mu(r)} - \frac{1}{\mathcal{A}}\text{,}
  \end{equation}
  illustrating that $q_\text{ion}$ is nearly constant throughout the atmosphere. As a consequence, $\Gamma_\text{e}$ 
  has only a very weak radial dependence.\footnote{Instead of using $q_\text{ion}$, one also
  finds the specific electron-scattering coefficient $s_\text{e} = n_\text{e} \sigma_\text{e} / \rho$
  in the literature \citep[e.g.,][]{Mihalas1978Book}.} 
  
  While Eq.\,(\ref{eq:Gammae}) is the common definition for
  $\Gamma$ in the context of the Eddington limit, it is only a fraction of the
  total outward radiative acceleration $a_\text{rad}$. Apart from the Thomson opacity 
  corresponding to scattering of free electrons, additional opacities originate in line 
  (bound-bound) and continua (bound-free, free-free) transitions, contributing to
  the total radiative acceleration.  $a_\text{rad}$ is thus separated into three parts:
  \begin{equation}  
   \label{eq:aradsep}
    a_\text{rad} = a_\text{thom} + a_\text{lines} + a_\text{true cont}
  .\end{equation}

  The last term $a_\text{true cont}$ refers to the acceleration originating from
  bound-free and free-free continuum transitions. As the electron opacity also forms 
  a continuum, these two terms are combined and referred to as the ``continuum''.
  In contrast, the term without electron scattering is sometimes called 
  the ``true continuum''. To avoid any confusion, we will adopt this notation in 
  the following.
   
  Similar  to Eq.\,(\ref{eq:Gammae}), we can thus define a
  $\Gamma,$ which corresponds to the total acceleration, 
  \begin{align} 
  \label{eq:Gammarad}
    \Gamma_\text{rad}(r) & := \frac{a_\text{rad}(r)}{g(r)} = \frac{1}{g(r)}\frac{4 \pi}{c} \frac{1}{\rho(r)} \int\limits_0^\infty\,\kappa_\nu(r)\,H_\nu(r)\,\mathrm{d}\nu,  
  \end{align}
  where $\kappa_\nu$ and $H_\nu$ are the opacity and Eddington flux at the
  frequency $\nu$, respectively, and $c$ is the speed of light.
  This ``full'' $\Gamma_\text{rad}$ is then used in the quasi-hydrostatic
  domain. A similar approach was also adopted 
  by \citet{LH2003} in their plane-parallel TLUSTY code, which is widely used 
  for calculating photospheric spectra of static atmospheres.
  
\begin{table*}%
        \caption{Characteristics of different stellar atmosphere codes concerning their treatment of the (quasi-)hydrostatic regime}
        \label{tab:codecmp}
        \centering
        \begin{tabular}{p{0.18\textwidth}p{0.19\textwidth}p{0.18\textwidth}p{0.16\textwidth}p{0.16\textwidth}}
        \hline\hline
           \rule[0mm]{0mm}{3mm}  & \textsc{PoWR}  &  \textsc{CMFGEN}  &  \textsc{FASTWIND}  &  \textsc{TLUSTY}  \\
        \hline
     \rule[0mm]{0mm}{4mm}\noindent
             radiative transfer  & comoving frame &  comoving frame   &  CMF/Sobolev\tablefootmark{a} &  static (obs. frame) \\
     \rule[0mm]{0mm}{4mm}\noindent
             blanketing          &     full       &     full          &   approximative     &   full            \\
     \rule[0mm]{0mm}{4mm}\noindent
             temperature stratification obtained by   & radiative equilibrium\tablefootmark{b} or thermal balance  &   radiative equilibrium\tablefootmark{b,c}  &   thermal balance   &   radiative equilibrium\tablefootmark{b} \\
     \rule[0mm]{0mm}{4mm}\noindent
             photosphere            & quasi-hydrostatic & quasi-hydrostatic & quasi-hydrostatic & hydrostatic \\
     \rule[0mm]{0mm}{4mm}\noindent
          $(\rho,\varv)$-update  & consistent & injections\tablefootmark{d} & start iterations\tablefootmark{e} & $\varv\,\equiv\,0$, $\rho$ consistent\tablefootmark{f} \\
     \rule[0mm]{0mm}{4mm}\noindent
          radiative acceleration used in hydrostatic eq. &  full $a_\text{rad}$ & full $a_\text{rad}$ & $a_\text{cont}$ approximation\tablefootmark{g} & full $a_\text{rad}$ \\
     \rule[0mm]{0mm}{4mm}\noindent
          $\varv_\text{turb}$ parameter in hydrostatic eq.? & yes  & yes & no & yes \\
     \rule[0mm]{0mm}{4mm}\noindent
          domain connection      & continuous $\mathrm{d}\varv/\mathrm{d}r$\tablefootmark{h} & $0.75\,\varv_\text{sonic}$\tablefootmark{i} & $0.1\,\varv_\text{sonic}$ (adjustable) &  no wind domain \\
                \hline
        \end{tabular}
  \tablefoot{
        \tablefoottext{a}{Comoving frame (CMF) method for main elements, Sobolev approximation for trace elements. Details are given in \citet{Puls+2005}.}
        \tablefoottext{b}{In PoWR and TLUSTY, the radiative equilibrium is considered in two different flavors, the so-called integral form, $4\pi \int \kappa_\nu \left(S_\nu - J_\nu\right) \mathrm{d}\nu = 0,$ and the flux consistency $4 \pi \int H_\nu \mathrm{d}\nu = \sigma_\text{SB} T_\text{eff}^4$ \citep{HL1995,HG2003}. CMFGEN instead only uses the integral form \citep{H2003}.}
        \tablefoottext{c}{In CMFGEN, the thermal balance is called ``electron energy balance'' (EEB) and used to check convergence and superlevel assignments, but not explicitly for temperature corrections \citep{HM1998,H2003}.}
        \tablefoottext{d}{After convergence of the statistical equations, the stratification in the quasi-hydrostatic part is adjusted and the model calculation is ``restarted'' from a gray temperature distribution. The total number of these ``restarts'', which are also referred to as ``injections'', must be specified beforehand.}
        \tablefoottext{e}{At the start of a model an iteration is done for the quasi-hydrostatic domain, where a Rosseland optical depth based on LTE opacities is calculated along with the calculation of the temperature and density stratification. The details are explained in \citet{SPH1997}.}
        \tablefoottext{f}{The hydrostatic equation is part of the set of linearized equations, which are solved consistently in every iteration \citep{Hubeny+1995}.}
        \tablefoottext{g}{The continuum acceleration is approximated by a nonintegral term with a parameterized Rosseland opacity \citep{SPH1997}.}
        \tablefoottext{h}{The continuous velocity gradient is the standard option in PoWR to find a connection point. Alternatively,
        the user can specify that the connection point is forced at $\varv = f \cdot \varv_\text{sonic}$.}
        \tablefoottext{i}{Models before $\sim$2013 used $\varv = 0.5\,\varv_\text{sonic}$ for the connection point \citep{Massey+2013}.}
  }  
\end{table*}

  
  In contrast to $\Gamma_\text{e}$, which is already known  at the start of a non-LTE model atmosphere calculation,  with the exception of the
exact value
  of $q_\text{ion}(r),$   the 
  full $\Gamma_\text{rad}$ has to be calculated iteratively. The radiative acceleration 
  is calculated from the population
  numbers, which in turn depend on the radiation field and also on the density 
  which itself is connected to $\Gamma_\text{rad}$ via the hydrostatic equation.
  To ensure a consistent solution,  
  the velocity field in the quasi-hydrostatic domain is constantly updated as soon 
  as the hydrostatic equation is violated by more than 5\%. This turned
  out to be sufficient in all test cases, as long as $\Gamma_\text{rad}(r) < 1$
  in the quasi-hydrostatic part. In reality, values larger than unity in the (quasi-)hydrostatic domain 
  would reflect additional physical phenomena occurring in the photosphere 
  \cite[e.g.,\ subphotospheric convection,][]{Cantiello+2009},
  which are not handled by our model atmospheres. The plane-parallel TLUSTY code
  treats only values of $\Gamma_\text{rad} \leq 0.9$ in the hydrostatic equation
  \citep{LH2003} as the hydrostatic equation fails for values larger than unity and
  numerical instabilities already occur when $\Gamma_\text{rad}$ gets close to this limit. We
  use a similar approach in PoWR for the quasi-hydrostatic equation, i.e., the value of $\Gamma_\text{rad}$ 
  is limited to $0.9$ for our velocity calculations. 

  \subsection{The quasi-hydrostatic treatment in different stellar atmosphere codes}
    \label{sec:codecmp}

  A couple of non-LTE stellar atmopshere codes are used in
  the field of hot and massive stars. Widely used are CMFGEN \citep{HM1998},
  FASTWIND \citep{SPH1997}, and, in case of negligible stellar winds, TLUSTY \citep{Hubeny+1995}.
  While these codes have a lot of similarities in their concepts, they also differ in certain 
  approaches, such as their treatment of the radiative transfer or their superlevel approach.
  Recent comparisons between CMFGEN and FASTWIND \citep{Massey+2013} for particular objects have 
  shown a good agreement in the derived temperatures, but also revealed that the surface
  gravities derived with FASTWIND are abound $0.12\,$dex lower than with CMFGEN. As the derived
  surface gravities are directly connected to the treatment of the quasi-hydrostatic layers,
  it is of major interest to understand the difference between the codes in this regime.
  
  Table \ref{tab:codecmp} provides an overview of the features affecting the quasi-hydrostatic treatment 
  of PoWR, CMFGEN, and FASTWIND together with the completely hydrostatic treatment in TLUSTY.
  The latter provides an interesting test case in the limit of negligible mass-loss rates, which
  will be further discussed in Sect.\,\ref{sec:tlusty}, where we compare spectra from TLUSTY
  models with those from PoWR. In CMFGEN, TLUSTY model atmospheres can also be used as a
  start approach. 
  
  The most striking difference in the quasi-hydrostatic treatment between the different wind
  codes (PoWR, CMFGEN, FASTWIND) is the update mechanism of the density and velocity stratification:
  
  \begin{itemize}
    \item In PoWR, we update the velocity field as described in Sect.\,\ref{sec:physics} as part
       of the main iteration as soon as the hydrostatic equation is violated by more than 5\% or
       the specified value of $\tau_\text{max}$ is missed by more than a specified $\epsilon_\tau$. 
       For each update, $\Gamma_\text{rad}(r)$ is applied in the hydrostatic equation, using
       $a_\text{rad}(r)$ from the comoving frame (CMF) calculations. With the new velocity field given,
       the stratification is updated according to the equation of continuity (\ref{eq:cont}).
    \item In CMFGEN \citep{HM1998}, the hydrostatic domain is first taken from a TLUSTY model or
       an old CMFGEN model, but can be updated with regards to the hydrostatic equation after a 
       first convergence of the population numbers and the radiation field. After each such stratification 
       update, sometimes referred to as ``injection'', the model iteration has to be continued  to reachieve convergence \citep[see, e.g.,][for a brief description]{Martins+2012}. 
       Similar to PoWR, the radiative acceleration used in the hydrostatic equation is taken from 
       the CMF radiative transfer calculations.
    \item In FASTWIND \citep{SPH1997},  the line contributions are instead neglected in the quasi-hydrostatic 
       domain and uses an approximate handling for the continuum acceleration, including both Thomson 
       and true continuum term. This simplification allows FASTWIND to set up the quasi-hydrostatic 
       domain in an iteration right at the start of a model,  in line with the goal of the code 
       of  being significantly faster than PoWR and CMFGEN. 
  \end{itemize}

  Furthermore, all three wind codes use a different criterion for the 
  connection point between the quasi-hydrostatic and the wind domain. While FASTWIND and CMFGEN use a 
  fixed fraction of the sound speed to define this connection point, PoWR puts the condition of a smooth
  velocity gradient between the two regimes. 

  Instead of treating only the quasi-hydrostatic part consistenly, one could also
  aim at a hydrodynamically consistent treatment throughout the whole model atmosphere, 
  i.e., including the wind domain. These calculations are typically not included in 
  state-of-the-art stellar atmosphere codes, especially as a prescribed $\beta$-law provides
  a good approximation for the outer wind regime of OB stars. A self-consistent calculation of
  the velocity field in the whole stellar atmosphere has been implemented as a nonstandard 
  option in PoWR in order to discuss WR winds \citep{GH2005,GH2008}.
  However, such models require much longer computation times and this option is not 
  required for the purpose, in which we focus on the quasi-hydrostatic regime. 
  The implementation presented here is completely sufficient in achieving a self-consistent
  stratification of the quasi-hydrostatic layers.
  Nevertheless, a revised implementation of completely hydrodynamically consistent 
  model atmospheres in the PoWR code will be addressed in detail in a future work.

\subsection{The effective gravity}
  \label{sec:geff}    
  
  The hydrostatic Eq.\,(\ref{eq:hydrostat}) can be written in an even
  shorter form with the definition of the \textit{effective gravity,}  
  \begin{equation}
    \label{eq:geffdef}
    g_\text{eff}(r) := \frac{G M_\ast}{r^2} \left[ 1 - \Gamma(r) \right]\text{.}
  \end{equation}
  To avoid any confusion, we will use $g_\text{grav}(r)$ instead of $g(r)$
  with $g_\ast = g_\text{grav}(R_\ast)$ to denote the full gravity from here on. 
  The effective gravity $g_\text{eff}$ is a fundamental fitting parameter for O- and
  B-stars when comparing models with observations. The pure gravitational acceleration
  $g_\text{grav}(r) = G M_\ast r^{-2}$, from which the stellar mass is derived, can only
  be calculated if $\Gamma_\text{rad}$ is known. Therefore, $g_\text{grav}$ is model 
  dependent. In these models an accurate treatment of the complete radiative pressure is required. 
  Correcting the observed $g_\text{eff}$ for Thomson pressure $\Gamma_\text{e}$ only 
  neglects a large part of the radiative pressure, and leads to an underestimated stellar 
  mass, as we will demonstrate in Sect.\,\ref{sec:results}.
  In the cases of negligible radiative pressures only do $\log g_\text{grav}$ and $\log g_\text{eff}$ 
  become indistinguishable. Otherwise, spectroscopic masses are 
  underestimated if calculated directly via $R_\ast^2\, g_\text{eff} / G$. 
  
  As the shape of a spectral line like H$\delta$ depends on $g_\text{eff}$ and not
  $g_\text{grav}$, there is a parameter degeneracy. A star with a weaker radiative
  pressure will produce the same line profiles as a star with a higher radiative pressure
  and a higher gravity.

  One of the major reasons for a strong model dependency of the stellar 
  mass is the large impact of the opacity on the radiative pressure. It 
  is significant which opacities are taken into account in the model. 
  Neglecting elements such as Fe or Ne will result in a lower $\Gamma_\text{rad}$ 
  and is not legitimate, even if the corresponding element is not under consideration
  in the intended analyses. The inferred stellar mass therefore depends not only on the
  adopted stellar parameters, but also on the details of the model atoms. 
  
  The adopted Doppler broadening velocity $\varv_\text{dop}$ used in the 
  comoving-frame calculations also has a significant impact: larger 
  velocities allow atoms to absorb photons in a wider frequency range, and 
  thus generally increase the radiative pressure in the atmosphere.

  Even though $g_\text{eff}(r)$ and $\Gamma_\text{rad}(r)$ are calculated 
  for each depth point in the PoWR code  to fulfill the hydrostatic equation, 
  there is a need for a certain ``reference value'' , which can be used for
  comparisons. As the emergent spectrum is mainly formed at $\tau_\text{Ross} \approx 2/3$ 
  in the photosphere, the first idea would be to use $g_\text{eff}$ at this value. 
  However, such a fixed point might already be located in the wind. 
  Therefore we define a weighted mean of $\Gamma_\text{rad}$ as
  \begin{equation}
    \label{eq:gammaradmean}
    \overline{\Gamma}_\text{rad} := \int\limits_{\tau_\text{max}}^{\tau_\text{sonic}} \Gamma_\text{rad}(\tau_\text{Ross})~e^{-\tau_\text{Ross}}~\mathrm{d}\tau_\text{Ross}\text{.}
  \end{equation}
  The upper limit of the integral $\tau_\text{sonic}$ usually denotes the optical depth of the 
  sonic point. However, we do not allow that this value drops below $0.1$, even in the rare cases
  where the actual sonic point would be in such a regime. The calculated value of
  $\overline{\Gamma}_\text{rad}$ is used to relate 
  $\log g_\text{eff}$ and $\log g_\ast = \log g_\text{grav}(R_\ast)$ via
 \begin{equation}
    \label{eq:geffggrav}
  \log g_\text{eff} = \log g_\ast + \log \left(1 - \overline{\Gamma}_\text{rad}\right).
 \end{equation}  

\begin{figure}[ht]
  \resizebox{\hsize}{!}{\includegraphics[angle=0]{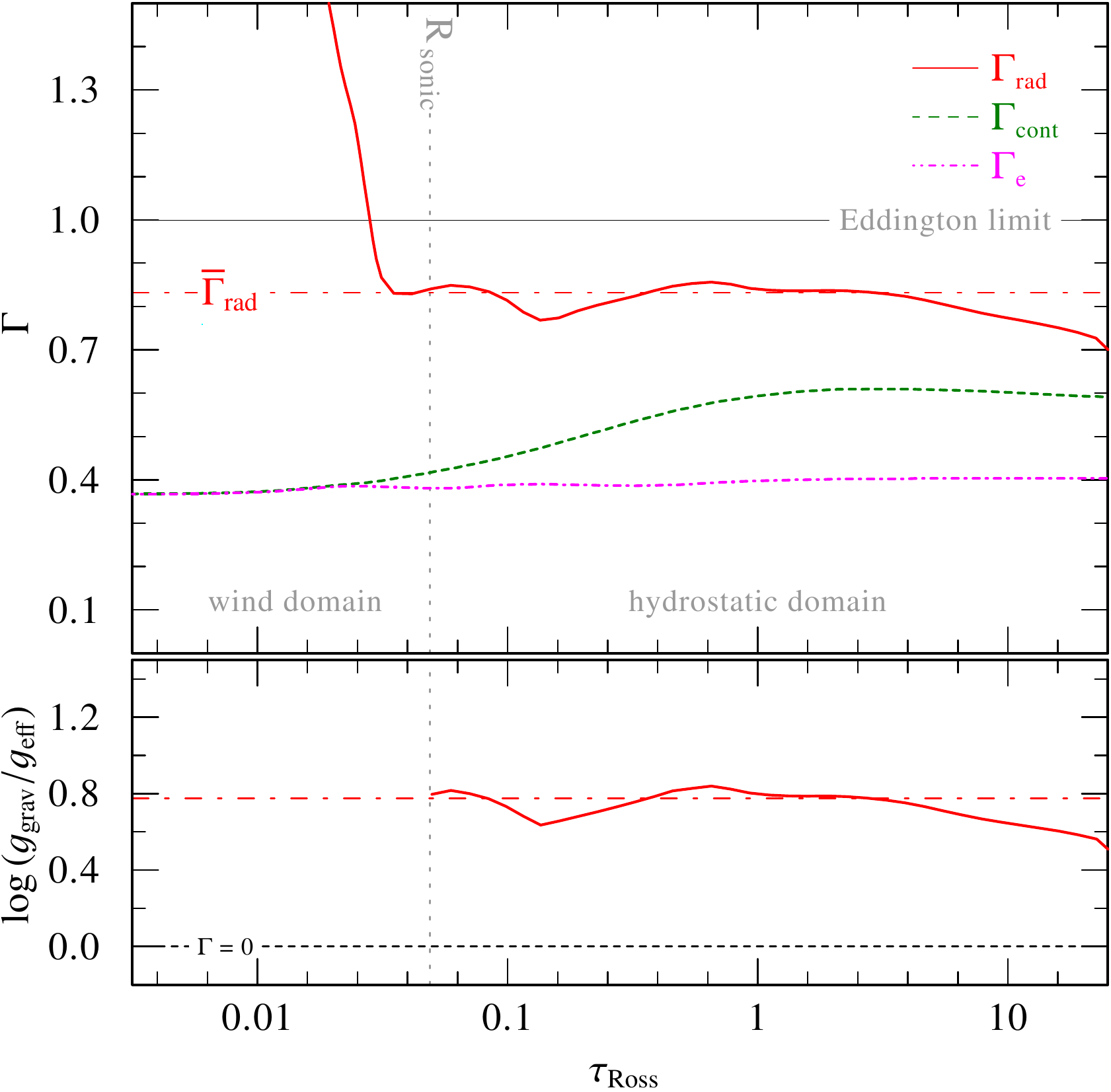}}
  \caption{Radial dependence of $\Gamma_\text{rad}$ (red, solid line, upper panel) for
           our supergiant test model, plotted 
           over the Rosseland optical depth $\tau_\text{Ross}$. The mean value obtained with 
           Eq.\,(\ref{eq:gammaradmean}) is indicated as a dashed-dotted line. For comparison,
           $\Gamma_\text{e}$ and the total continuum contribution $\Gamma_\text{cont}$ are
           also shown. In the lower panel, the difference between $g_\text{grav}$ and
           $g_\text{eff}$ is illustrated: the red solid curve marks the actual difference at
           each optical depth, while the dashed-dotted line denotes the difference using
           $\overline{\Gamma}_\text{rad}$. The gray dashed vertical line denotes the location
           of the sonic point.}
  \label{fig:gammarad}
\end{figure}
  The radial dependence of $\Gamma_\text{rad}$ for a supergiant test model 
  is illustrated in Fig.\,\ref{fig:gammarad},
  where it is plotted against the total Rosseland opacity. It is evident that in the
  quasi-hydrostatic regime not only the line transitions, but also the atomic
  continuum transitions cannot be neglected, while this quickly changes as soon as we enter the 
  supersonic domain, where $\Gamma_\text{cont}$ consists only of $\Gamma_\text{e}$.
  One can see that the calculated 
  value of $\overline{\Gamma}_\text{rad}$ is indeed representative for the difference between
  $g_\text{grav}$ and $g_\text{eff}$ in the photosphere. 
  Values of $\overline{\Gamma}_\text{rad}$ for a series of test models are
  shown and discussed in Sect.\,\ref{sec:results}.
 
 
  The reference radius for all values of $g_\text{eff}$ given in this work is $R_\ast$.
  However,  other radii are also used in the literature, typically 
  $R_\frac{2}{3} = R\left(\tau_\text{Ross} = \frac{2}{3}\right)$ or the radius of the
  sonic point $R_\text{s}$. The latter roughly indicates the outer end of the quasi-hydrostatic domain.
  It is therefore interesting to check how much the values of $g_\text{eff}$ and $g_\text{grav}$
  are affected when referring to different radii. As this effect is largest for supergiants,
  we examine an O-star model with $T_\ast = 32.5\,$kK, $\log L/L_\odot = 5.6$,
  $\log g_\text{grav}(R_\ast) = 3.25$ and $\log \dot{M} = -5.75\,$[M$_\odot$/yr]. For this model,
  we have $R_\frac{2}{3} = 1.06\,R_\ast$ and the sonic point is located at $R_\text{s} = 1.11\,R_\ast$,
  leading to differences of $0.05\,$dex and $0.09\,$dex in $\log g$. Especially the latter value is
  approximately what can be achieved by accurate fitting of high quality spectra and, therefore, we have to take
  the reference radius  into account when analyzing these objects. 
  This effect is lower for giants and dwarfs, namely on the order
  of $0.03\,$dex and $0.01\,$dex for the sonic point, respectively.
  

   In the literature, the term $g_\text{eff}$ may not always have the same meaning.
   Especially when dealing with observations, it is often not clearly stated whether
   values termed as $\log g$ have been corrected for radiative acceleration. Values
   that have been corrected for centrifugal acceleration in a statistical sense are sometimes labeled
   as ``true'' gravities $g_\text{true}$ or $g_\text{c}$ with the relation
   \begin{equation}
     \label{eq:centricorr}
     g_\text{c} = g + \frac{\left(\varv \sin i\right)^2}{R_{\ast}}
   \end{equation}
   \citep[][]{Repolust+2004}.
   The term labeled as $g$ in this equation
   is sometimes called $g_\text{eff}$ \citep[e.g.,][]{Massey+2013}. We would like
   to stress that this is not identical to our definition of
   the effective gravity in this work. The variable $g$ in (\ref{eq:centricorr}) refers
   to the gravitational acceleration specified in a non-rotating stellar atmosphere model, which
   has not been reduced by radiative acceleration. As such, $g$ would be referred to 
   as $g_\text{grav}$, following the notation of this work. 
   Caution is therefore adviced when dealing with the term ``effective gravity'', as 
   its definition might differ significantly between different authors.

  \section{Results and discussion}
    \label{sec:results}
    
        \subsection{Test model details}
        \label{sec:testmodels}

   To illustrate the impact of accounting for the full radiative pressure consistently,
   we calculate a set of PoWR models for a fixed temperature of $T_\ast = 32.5\,$kK
,   which corresponds to a late O-star. For this temperature, we calculate 
   three types of models with different surface gravities, corresponding
   to the luminosity classes I (supergiant), III (giant), and V (dwarf). As
   we focus on the quasi-hydrostatic part, we adopt only schematic parameters for the wind,
   i.e., all models have the same terminal wind velocity
   of $\varv_\infty = 2000\,$km/s. The outer model boundary is set to $100\,R_*$. 
   The velocity field in the wind domain can be 
   prescribed by a $\beta-$law with $\beta = 0.8$ (see Eq.\,\ref{eq:beta}), which is
   accurate enough for our purposes  as we do not want to analyze the outer wind. The 
   luminosities are ``typical'' representatives of their class according to \citet{MSH2005}. 
   We chose the surface gravities  similarly, but with slight adjustments of
   up to $0.1\,$dex, corresponding to the closest grid point in the 
   TLUSTY O-star model grid \citep{LH2003} for a later comparison 
   (see Sect.\,\ref{sec:tlusty}). We adopted mass-loss rates  from 
   \citet{VdKL2000}, and the wind is assumed to be smooth 
   \citep[no density contrast, i.e., $D = 1$ in the notation of][]{HK1998}.
   The chemical compositions
   are taken to be solar. Here, we use the abundances inferred by \citet{GS1998} 
   instead of the
   newer ones obtained by \citet{Asplund+2009} to have identical values to those used in
   the TLUSTY O-grid.
   All parameters of the 
   models are compiled in Table\,\ref{tab:testmodels}.   

\begin{figure}[ht]
  \resizebox{\hsize}{!}{\includegraphics[angle=270]{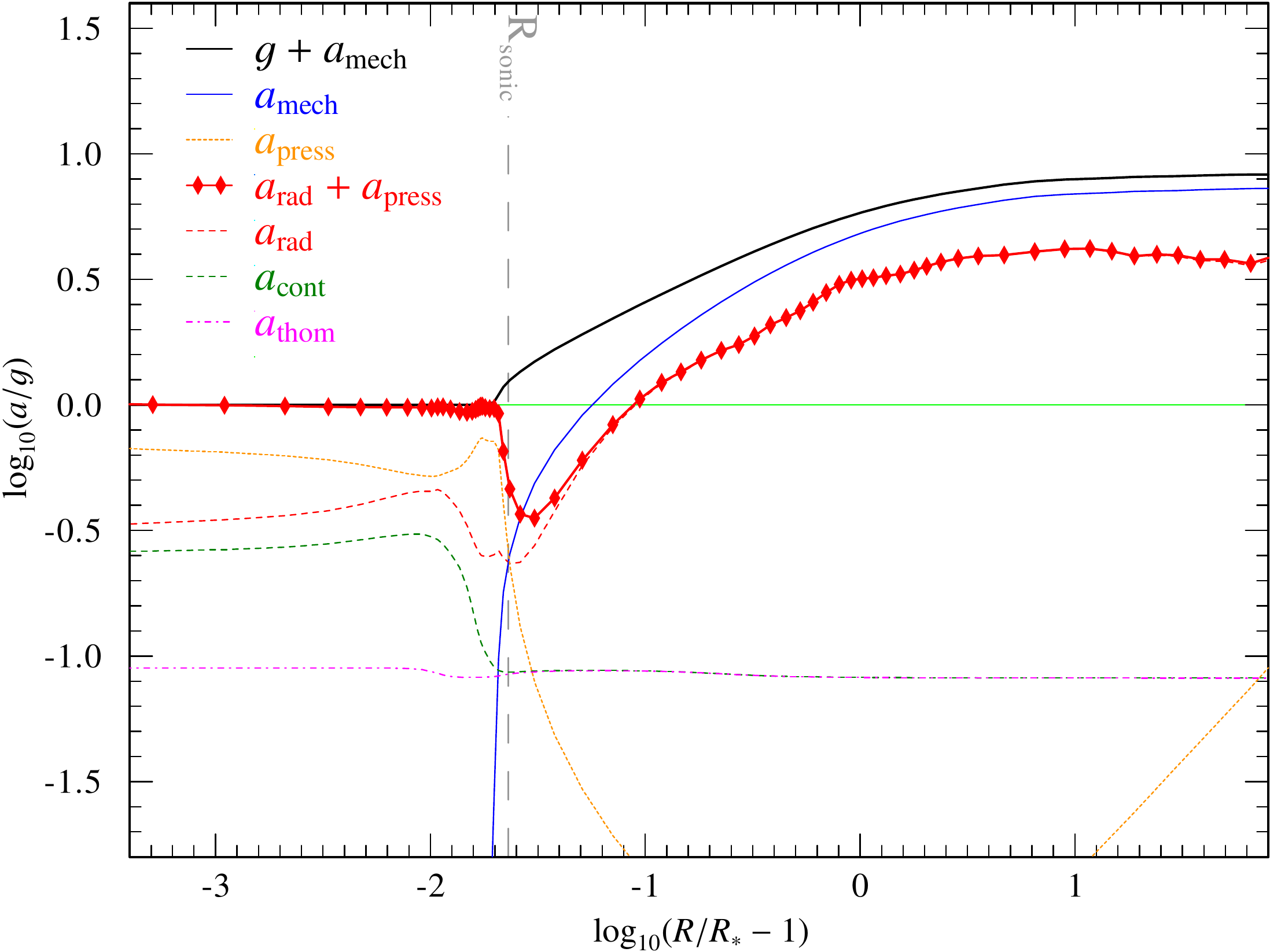}}
  \caption{Acceleration stratification for a model with a consistent quasi-hydrostatic
           domain (Type (b) models). The wind acceleration (thick red diamond line) is compared to
           the repulsive sum of inertia and gravitational acceleration $g(r)$ (black
           line). All terms have been normalized to $g(r)$.}
  \label{fig:hystfullstrat}
\end{figure}

\begin{figure}[ht]
  \resizebox{\hsize}{!}{\includegraphics[angle=270]{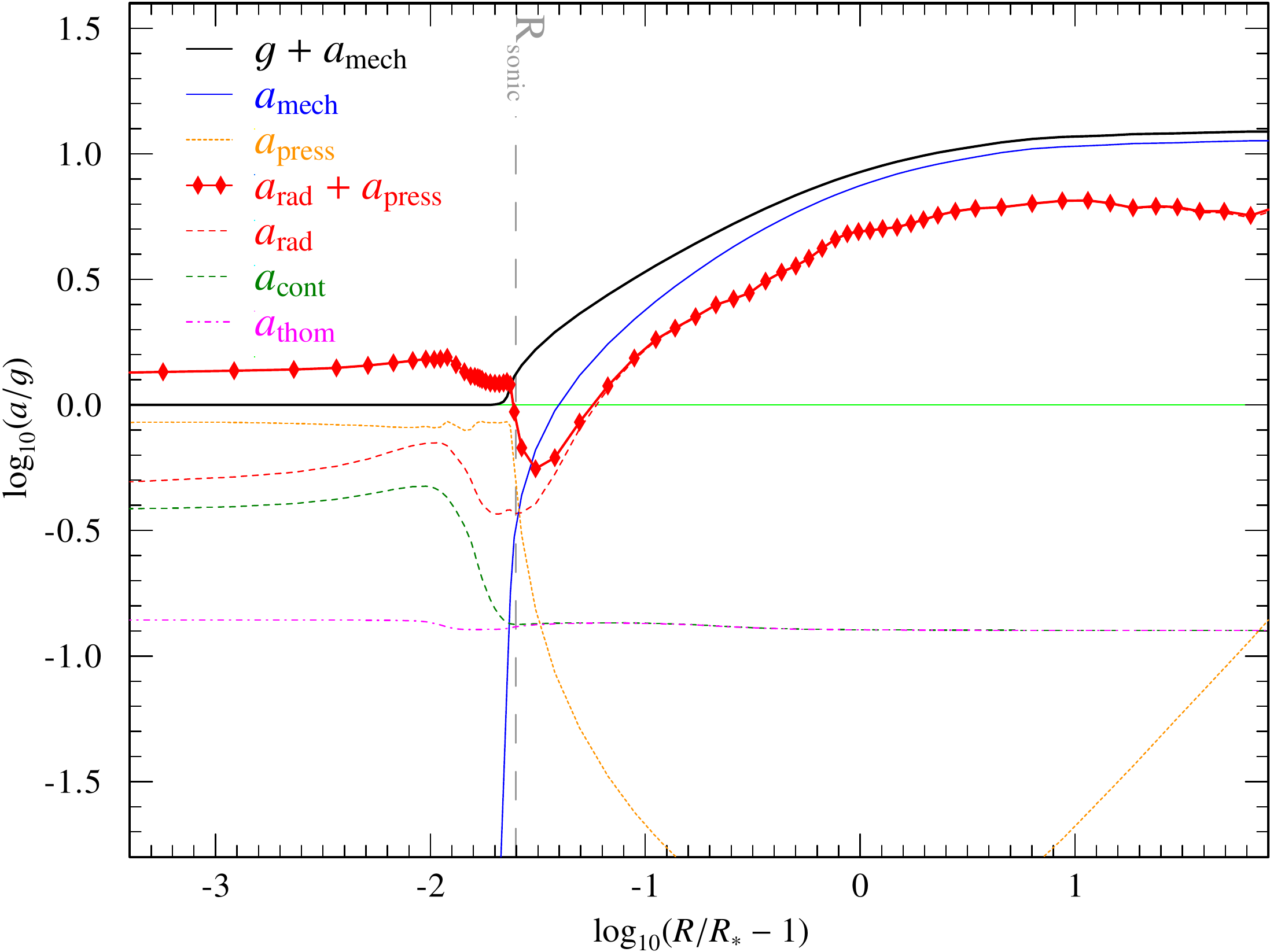}}
  \caption{Same as Fig.\,\ref{fig:hystfullstrat}, except  with a velocity law in the
           quasi-hydrostatic domain accounting only for $\Gamma_\text{e}$ (Type (a) models). While
           the wind domain is not affected, the acceleration balance in the
           quasi-hydrostatic part is significantly different.}
  \label{fig:hystestrat}
\end{figure}
    
\begin{table}
  \caption{O-star test model parameters}
  \label{tab:testmodels}
  \centering
  \begin{tabular}{l c c c}
  \hline\hline
    Luminosity class  \rule[0mm]{0mm}{3mm}                &   I     &   III   &    V     \\
  \hline
    $T_*$\,[kK]  \rule[0mm]{0mm}{3mm}                     & \multicolumn{3}{c}{ $32.5$ } \\
    $\log g_\text{eff}$\,[cm\,s$^{-2}$]\tablefootmark{a}  &  $2.47$ &  $3.02$ &  $3.79$  \\
    $\log g_\text{grav}$\,[cm\,s$^{-2}$]\tablefootmark{a} &  $3.25$ &  $3.50$ &  $4.00$  \\
    $\overline{\Gamma}_\text{rad}$                             &  $0.83$ &  $0.67$ &  $0.39$  \\
    $\Gamma_\text{e}$                                     &  $0.40$ &  $0.22$ &  $0.07$  \\
    $R_*$\,[$R_\odot$]                                    & $20.0$  & $14.1$  &  $8.1$   \\
    $\log \dot{M}$\,[$M_\odot\,\text{yr}^{-1}$]           & $-5.75$ & $-6.25$ & $-7.1$   \\
    $\log L$\,[$L_\odot$]                                 &  $5.60$ &  $5.30$ &  $4.82$  \\
    $M_\ast$\,[$M_\odot$]                                 & $25.9$  & $23.0$  & $24.1$   \\
    $\beta$                                               & \multicolumn{3}{c}{0.8}      \\                 
    $\varv_\infty$\,[km/s]                                & \multicolumn{3}{c}{2000}     \\                 
    $\varv_\text{turb}$\,[km/s]                           & \multicolumn{3}{c}{10}       \\                 
    \medskip
    $\varv_\text{dop}$\,[km/s]                            & \multicolumn{3}{c}{20}       \\                 
    $X_\text{H}$\tablefootmark{b} \rule[0mm]{0mm}{3mm}    & \multicolumn{3}{c}{$0.704$}  \\
    $X_\text{He}$\tablefootmark{b}                        & \multicolumn{3}{c}{$0.282$}  \\
    $X_\text{C}$\tablefootmark{b}                         & \multicolumn{3}{c}{$2.78 \times 10^{-3}$}  \\
    $X_\text{N}$\tablefootmark{b}                         & \multicolumn{3}{c}{$8.14 \times 10^{-4}$} \\
    $X_\text{O}$\tablefootmark{b}                         & \multicolumn{3}{c}{$7.56 \times 10^{-3}$} \\
    $X_\text{Mg}$\tablefootmark{b}                        & \multicolumn{3}{c}{$6.45 \times 10^{-4}$} \\
    $X_\text{Al}$\tablefootmark{b}                        & \multicolumn{3}{c}{$5.56 \times 10^{-5}$} \\
    $X_\text{Si}$\tablefootmark{b}                        & \multicolumn{3}{c}{$6.96 \times 10^{-4}$} \\
    $X_\text{P}$\tablefootmark{b}                         & \multicolumn{3}{c}{$6.10 \times 10^{-6}$} \\
    $X_\text{S}$\tablefootmark{b}                         & \multicolumn{3}{c}{$4.79 \times 10^{-4}$} \\
    $X_\text{Fe}$\tablefootmark{b,c}                      & \multicolumn{3}{c}{$1.34 \times 10^{-3}$} \\
  \hline
  \end{tabular}
  \tablefoot{
        \tablefoottext{a}{Values refer to the consistent models. Comparison model results are given in Table\,\ref{tab:gcomp}.}
        \tablefoottext{b}{Solar abundances, as obtained by \citet{GS1998}, specified here as mass fractions.}
        \tablefoottext{c}{Fe includes also the further iron group elements Sc, Ti, V, Cr, Mn, Co, and Ni.
                          See \citet{GKH2002} for relative abundances.}
  }  
\end{table} 
  
\begin{figure*}[ht]
  \centering
    \includegraphics[angle=0,width=\textwidth]{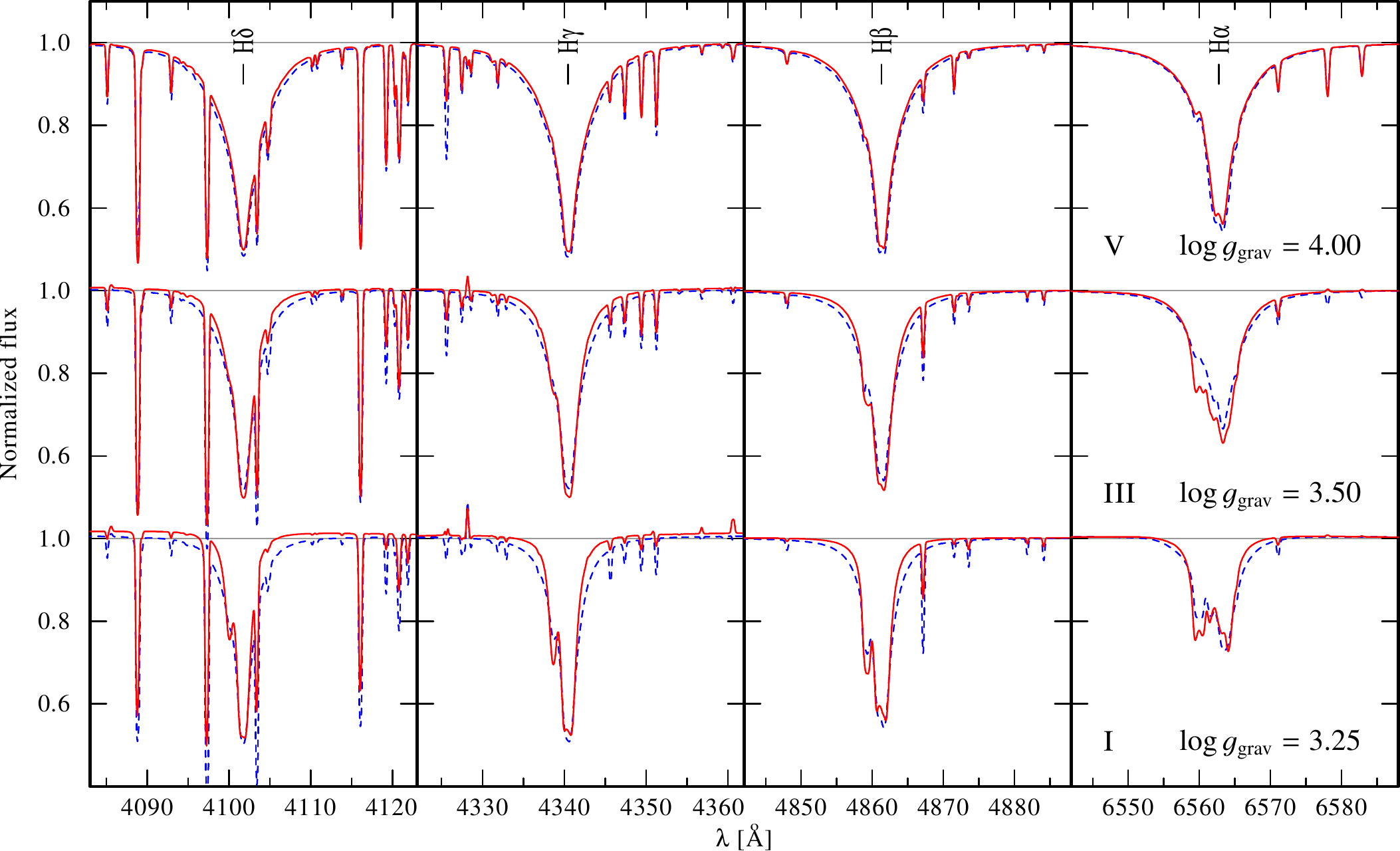}
  \caption{Balmer line profiles of H$\delta$ (left) to H$\alpha$ (right) 
           for test models with fixed $g_\text{grav}$. The spectra from 
           the (a)-type models with a quasi-hydrostatic stratification calculated using
           only the Thomson term $\Gamma_\text{e}$ are shown as blue dashed line. 
           The red solid
           lines display the profiles from the (b)-type models accounting for 
           the full radiative term $\Gamma_\text{rad}$.}
  \label{fig:gravfix-multi}
\end{figure*}
  
\begin{table}
  \caption{O-star test model type overview}
  \label{tab:testtypes}
  \centering
  \begin{tabular}{l c c c}
  \hline\hline
   Model type  \rule[0mm]{0mm}{3mm}      &   (a)                  &   (b)                          &    (c)   \\
  \hline  
   $g_\text{grav}$\rule[0mm]{0mm}{3.5mm} & fixed\tablefootmark{2} & fixed\tablefootmark{2}         &    calculated              \\
   $\Gamma$\tablefootmark{1}             & $\Gamma_\text{e}$      & $\overline{\Gamma}_\text{rad}$ &    $\Gamma_\text{e}$       \\
   $g_\text{eff}$                        & calculated             & calculated\tablefootmark{3}    &    fixed\tablefootmark{3}  \\
  \hline
  \end{tabular}
 \tablefoot{
        \tablefoottext{1}{$\Gamma$ value used to relate $g_\text{geff}$ and $g_\text{grav}$}
        \tablefoottext{2}{Type (a) and (b) models within the same luminosity class have the same $g_\text{grav}$}
        \tablefoottext{3}{A type (c) model has the same $g_\text{geff}$ as the type (b) model of the same luminosity class}
  }  
\end{table}
  
  For each of the three luminosity classes, we calculate three models, which we refer to as (a), (b), and (c) in the 
  following, making nine models in total period. 
  Models (a) and (b) have a fixed $g_\text{grav}$ , which corresponds to the respective luminosity class 
  and differ only in the treatment of the radiative acceleration 
  in the quasi-hydrostatic part: models (a) are only calculated with $\Gamma_\text{e}$ , while models (b) 
  include the full $\Gamma_\text{rad}$, i.e., they account for the complete radiative pressure. Typical
  examples are shown in Fig.\,\ref{fig:hystfullstrat} and Fig.\,\ref{fig:hystestrat}, where we plot
  the acceleration stratifications for a dwarf (class V) model. One can see that Fig.\,\ref{fig:hystfullstrat}
  shows the consistent (b)-model, where the sum of the total radiative acceleration and gas pressure equals gravitation
  in the quasi-hydrostatic domain. The result of the (a)-model is shown in Fig.\,\ref{fig:hystestrat}.
  Now only $\Gamma_\text{e}$ and gas pressure are taken into account for the hydrostatic equation and thus
  the sum of the total radiative acceleration and gas pressure is larger than the local gravity in the 
  subsonic part. The outer parts
  of both models are very similar as both use the same prescribed $\beta$-velocity law.
        
  In models (c), like in models (a), we account only for the Thomson term $\Gamma_\text{e}$ 
  in the quasi-hydrostatic domain, but instead of specifying $\log g_\text{grav}$, we specify the effective 
  gravity $\log g_\text{eff}$ and adopt its value from the corresponding model (b), calculated from Eq.\,(\ref{eq:geffggrav}). 
  An overview of the basic similarities and differences between all three model types is given in
  Table\,\ref{tab:testtypes}.
  Since models (b) and (c) of a given luminosity class have the same
  effective gravity, the wings of pressure-broadened lines are expected to be identical in both models.  
  However, because of  different treatment of the quasi-hydrostatic 
  domain, the actual surface gravities $g_\text{grav}$ and spectroscopic masses $M_\ast$
  implied from both models will differ. 

        \subsection{Comparison with fixed $g_\text{grav}$}
        \label{sec:ggravcomp}

  Figure\,\ref{fig:gravfix-multi} illustrates the impact of accounting for the full radiative 
  pressure on prominent Balmer lines (left to right: H$\delta$, H$\gamma$, H$\beta$, H$\alpha$)  
  for dwarfs (upper panels), giants (middle panels), and 
  supergiants (lower panels).
  The impact of either including only the Thomson term $\Gamma_\text{e}$ (model a, blue dashed lines) 
  or including the full radiative term $\Gamma_\text{rad}$ 
  (model b, red solid lines) can be seen in the line wings. Because of their 
  larger outward pressures, the quasi-hydrostatic domains of models (b) are less dense than 
  those of models (a), and, as a consequence, the line wings obtained by models (b) are narrower.
  For the same reason, the value of $g_\text{eff}$ is   
  always smaller in models (b) compared to models (a) (see below). 

  A further inspection
  of Fig.\,\ref{fig:gravfix-multi} reveals that the difference between models (a) and (b)  
  increases with luminosity class. A look at Eq.\,(\ref{eq:Gammarad}) reveals
  that lower values of $\log g_\text{grav}$ will increase $\Gamma_\text{rad}$, which is 
  indeed much larger for the supergiant than for the dwarf. Even though the difference between $\Gamma_\text{rad}$
  and $\Gamma_\text{e}$ is not getting much stronger with lower $\log g_\text{grav}$, as 
  $\Gamma_\text{e}$ also changes proportional to $R_\ast^2$,
  both values are significantly larger for the supergiant and thus the difference in the
  derived $g_\text{eff}$ value increases, which is reflected in the increasing 
  line wing difference in Fig.\,\ref{fig:gravfix-multi}.
  

  Even for our dwarf models 
  ($\log g_\text{grav} = 4.0$, upper panels), where the dashed and solid lines can only barely be
  distinguished, we obtain a difference of $\Delta \log g_\text{eff} = 0.21\,$dex when accounting 
  for the full radiative pressure via $\Gamma_\text{rad}$
  ($\log g_\text{eff} = 3.97$ for model (a) vs. $\log g_\text{eff} = 3.78$ for model (b)). 
  For our giant models ($\log g_\text{grav} = 3.5$, middle panels) 
  the difference increases to $0.37\,$dex ($3.39$ vs. $3.02$).
  Finally, for the supergiant models  
  ($\log g_\text{grav} = 3.25$, lower panels), a formidable
  difference of $0.56\,$\,dex 
  ($3.03$ vs. $2.47$) is obtained. Even though the impact on the spectral
  appearance of the hydrogen lines might not be particularly striking in Fig.\,\ref{fig:gravfix-multi}, 
  the differences in the corresponding values of $\log g_\text{eff}$ are quite remarkable. 

\begin{figure}[ht]
  \centering
    \resizebox{\hsize}{!}{\includegraphics[angle=0]{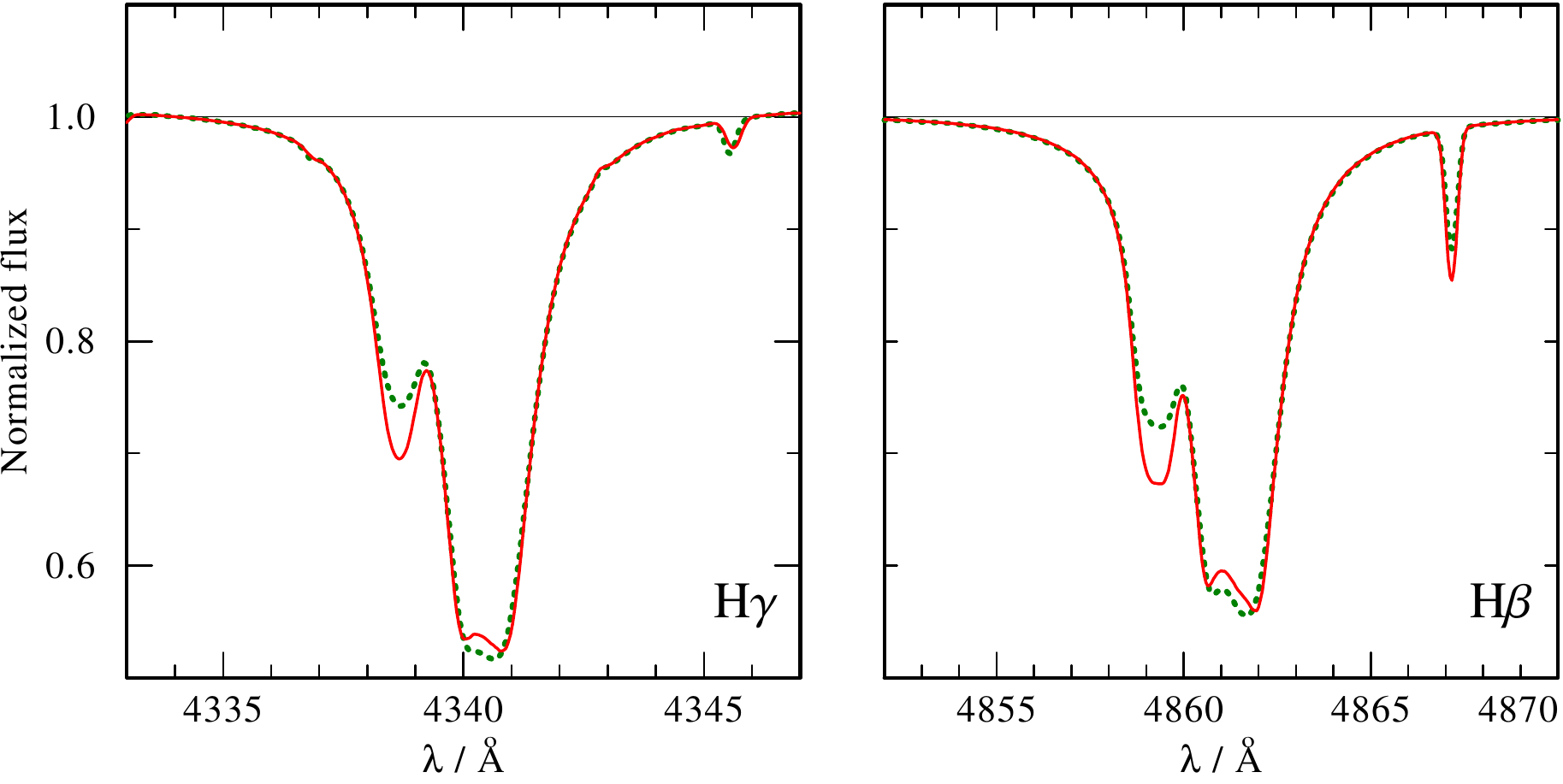}}
  \caption{The H$\gamma$ and H$\beta$ lines (left and right panels, respectively), as obtained
           from models (b) and (c) (red solid and green dotted lines, respectively), both 
           with an effective gravity of $\log g_\text{eff} = 2.47$. Although the strengh of the 
           line wings is the same, the two models imply masses and gravities, which strongly differ because of the different treatment of the radiative pressure (cf.\ Table\,\ref{tab:gcomp})}
  \label{fig:samegeff}
\end{figure}

        \subsection{Comparison with fixed $g_\text{eff}$}
        \label{sec:geffcomp}

  In the models (a) and (b) discussed above, we specify $g_\text{grav}$, $T_\ast$, and $L$, and
  therefore the radii, the spectroscopic 
  masses are known a priori. When dealing with observations, however, it is $g_\text{eff}$, and not $g_\text{grav}$, 
  which is empirically measurable. Thus, one would fix the  effective 
  gravity that best reproduces the observation, and let the actual gravity $g_\text{grav}$ 
  be a model output. Based on the detailed physics implemented 
  in the model atmosphere, the gravity $g_\text{grav}$ and the spectroscopic mass $M_\ast$ follow 
  from $\log g_\text{eff}$. This scenario is reflected in models (b) and (c), which are calculated with 
  identical effective gravities, but with a different treatment of the radiation pressure in the quasi-hydrostatic 
  domain. Fig.\,\ref{fig:samegeff} shows the 
  Balmer lines H$\gamma$ (left panel) and H$\beta$ (right panel) as obtained from the supergiant 
  models (b),, which account for $\Gamma_\text{rad}$ (red solid line), 
  and (c), which only include $\Gamma_\text{e}$  (green dotted line). As both models have the same $g_\text{eff}$,  
  the line wings obtained from both models can hardly be distinguished. However, some differences are 
  seen in the line cores and in the helium and metal lines. 
  
\begin{figure}[ht]
  \resizebox{\hsize}{!}{\includegraphics[angle=270]{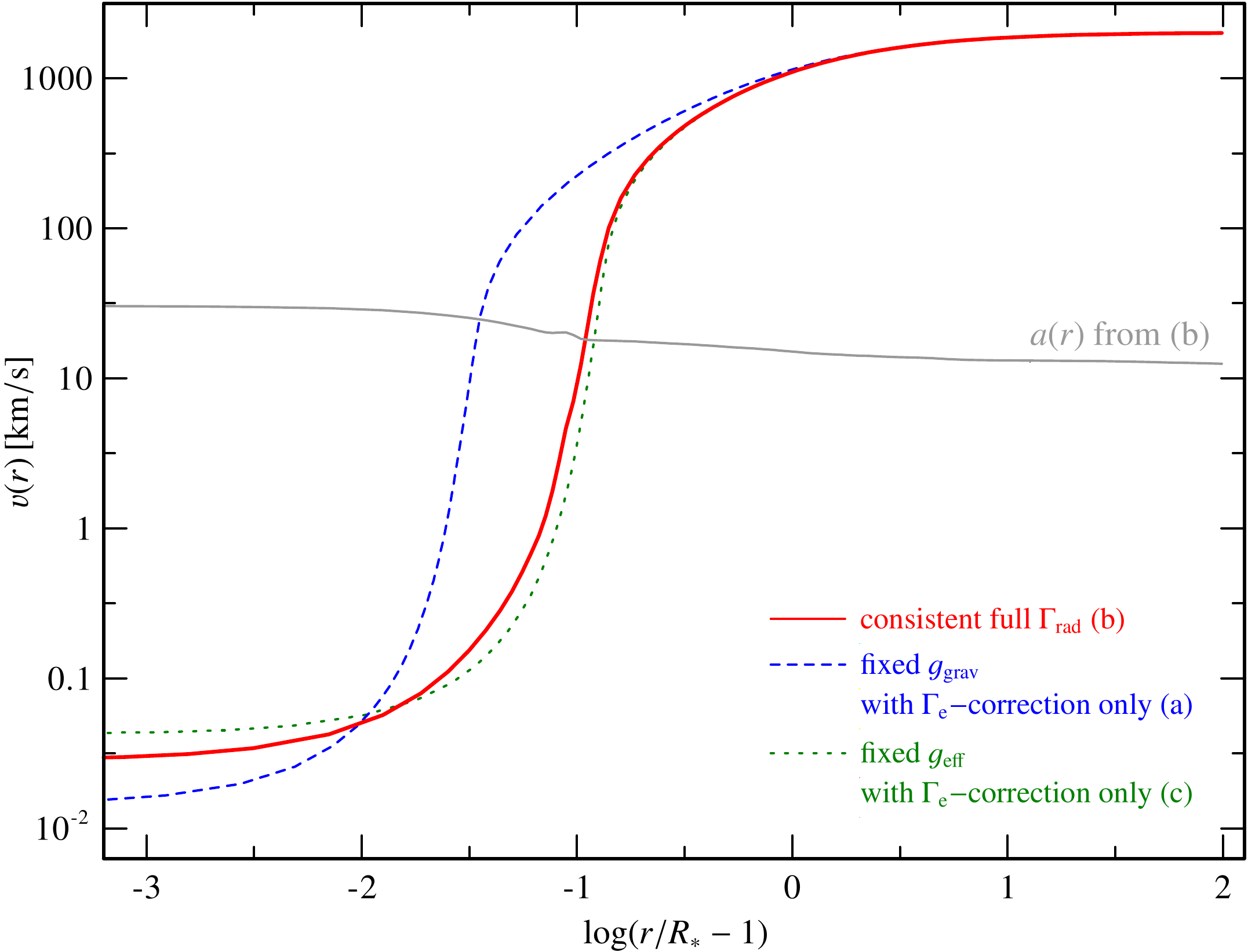}}
  \caption{Velocity stratification for three supergiant test models:
           the consistent quasi-hydrostatic model (b) (red solid) is compared to 
           a model (a) (blue dashed) with the same $g_\text{grav}$, but where only 
           $\Gamma_\text{e}$ is taken into account in the quasi-hydrostatic part (blue). 
           The third model (c) (green dotted) shares $g_\text{eff}$ with
           model (b), except it accounts only for $\Gamma_\text{e}$ and thus has
           a different $g_\text{grav}$.}
  \label{fig:vcmp}
\end{figure}

  To illustrate the effect of this different treatment, we show the velocity
  stratification for all three supergiant models in Fig.\,\ref{fig:vcmp}. All 
  models agree in the outer part where we have prescribed the $\beta$-law.
  However, in the inner part it becomes evident that fixing $g_\text{eff}$ indeed leads
  to relatively similar velocity stratifications with both the consistent
  $\overline{\Gamma}_\text{rad}$-approach and the 
  $\Gamma_\text{e}$-approach, which is reflected in the agreement of line profiles  
  between the models (b) and (c). If one fixes instead the more fundamental parameter $g_\text{grav}$,
  the resulting velocity stratification is significantly different in the inner part, if 
  accounting only for $\Gamma_\text{e}$, including a huge difference in the radial location
  of the sonic point. These discrepancies explain the spectral differences between 
  models (a) and (b) shown in Fig.\,\ref{fig:gravfix-multi}.

\begin{table}
  \caption{Deduced stellar masses with different quasi-hydrostatic approaches}
  \label{tab:gcomp}
  \centering
  \begin{tabular}{l c c c}
  \hline\hline
    Luminosity class  \rule[0mm]{0mm}{3mm}                                      &  I     &   III   &    V     \\
  \hline
    $\log g_\text{eff}$\,[cm\,s$^{-2}$]\tablefootmark{a} \rule[0mm]{0mm}{3.5mm} & $2.47$ &  $3.02$ &  $3.79$  \\
    $\log g_\text{grav}(\Gamma_\text{e})$\,[cm\,s$^{-2}$]\tablefootmark{b}      & $3.00$ &  $3.25$ &  $3.82$  \\
    $\log g_\text{grav}(\Gamma_\text{rad})$\,[cm\,s$^{-2}$]                     & $3.25$ &  $3.50$ &  $4.00$  \\
    $M_{\ast}(\Gamma_\text{e})$\,[M$_\odot$]\tablefootmark{b}                   & $14.5$ &  $12.9$ &  $16.2$  \\
    $M_{\ast}(\Gamma_\text{rad})$\,[M$_\odot$]                                  & $25.9$ &  $23.0$ &  $24.1$  \\
  \hline
  \end{tabular}
  \tablefoot{  
        \tablefoottext{a}{The effective gravity (cf. Sect.\,\ref{sec:geff}) is fixed for all models here}
        \tablefoottext{b}{These values are derived from models that were calculated as if only the Thomson radiative pressure would enter the hydrostatic equation.}
  }
\end{table}
  
  Despite the similarities between the models (b) and (c) seen in the line wings and in the
  velocity law, their values of $\log g_\text{grav}$ and $M_\ast$ differ significantly.
  Table\,\ref{tab:gcomp} shows the values of $\log g_\text{grav}$ and $M_\ast$ for the dwarf, 
  giant, and supergiant models. As could be anticipated from Eq.\,(\ref{eq:geffggrav}), the 
  models that account for the full radiative pressure (models b) have larger values for 
  $\log g_\text{grav}$ and $M_\ast$ because their $\Gamma,$ which is used to relate  
  $g_\text{eff}$ and $g_\text{grav}$ , is larger in these models. The spectroscopic
  masses deviate by roughly a factor of two. Interestingly, there is no clear trend for increasing deviation with 
  increasing luminosity class.
  Intuitively, one might expect that the deviation would have to be larger for the supergiant model,
  compared to the dwarf model, as we obtained it in spectral comparison of the models (a) and (b) which were 
  shown in Fig.\,\ref{fig:gravfix-multi}.
  However, the calculation of the (c) models is quantitatively different from the (a) models. Even though both
  consider only $\Gamma_\text{e}$ for obtaining the velocity field in the quasi-hydrostatic part, the (c) models
  are specifically designed to reproduce the $g_\text{eff}$-value obtained with the full $\Gamma_\text{rad}$ , while
  the (a) models have a completely different approach, with the $g_\text{grav}$-value being identical to the (b) models.
  In fact, we can estimate the masses of the (c) models with an easy calculation. Starting from the requirement
  that both, (b) and (c) models should have the same $g_\text{eff}$, it immediately follows via Eq.\,(\ref{eq:geffdef}) that  
  \begin{equation}
    \label{eq:bcmassexpl}
    M_\ast^{(c)} \left( 1 - \Gamma_\text{e}^{(c)} \right) = M_\ast^{(b)} \left( 1 - \overline{\Gamma}_\text{rad}^{(b)} \right)\text{.}
  \end{equation}
  The small superscripts indicate the value from the corresponding model family, i.e., $M_\ast^{(c)}$ is short for
  $M_\ast(\Gamma_\text{e})$ in the (c) models. As we do not know $\Gamma_\text{e}^{(c)}$ in advance, since it
  contains $M_\ast(\Gamma_\text{e})$ itself by definition (\ref{eq:Gammae}), we need to replace it with a value 
  from the (b) model. Because the luminosity $L$ is the same in the (b) and the (c) models and the ionization
  parameter $q_\text{ion}$ only changes marginally, we can deduce from Eq.\,(\ref{eq:Gammae}) that the product
  of $\Gamma_\text{e}$ and $M_\ast$ will be approximately the same for both models. Hence we get an
  expression that allows us to replace $\Gamma_\text{e}^{(c)}$ with the known $\Gamma_\text{e}^{(b)}$, i.e.,  \begin{equation}
    \Gamma_\text{e}^{(c)} \approx \frac{M_\ast^{(b)}}{M_\ast^{(c)}} \Gamma_\text{e}^{(b)}
  .\end{equation}
  Using that in (\ref{eq:bcmassexpl}) yields
  \begin{align}
    M_\ast^{(c)} - M_\ast^{(b)} \Gamma_\text{e}^{(b)} & \approx M_\ast^{(b)} \left( 1 - \overline{\Gamma}_\text{rad}^{(b)} \right) \\
    \label{eq:bcmassrel}
    M_\ast^{(c)} & \approx M_\ast^{(b)} \left[ 1 - \left( \overline{\Gamma}_\text{rad}^{(b)} - \Gamma_\text{e}^{(b)} \right) \right]\text{.}
  \end{align}
  
  This means that masses of the (c) models only depend  on the difference between $\overline{\Gamma}_\text{rad}$ 
  and $\Gamma_\text{e}$ in the consistent (b) models, not on the absolute values. The difference is comparable for
  all three luminosity classes, and so is the mass deviation in Table\,\ref{tab:gcomp}.

\begin{figure}[ht]
  \resizebox{\hsize}{!}{\includegraphics[angle=0]{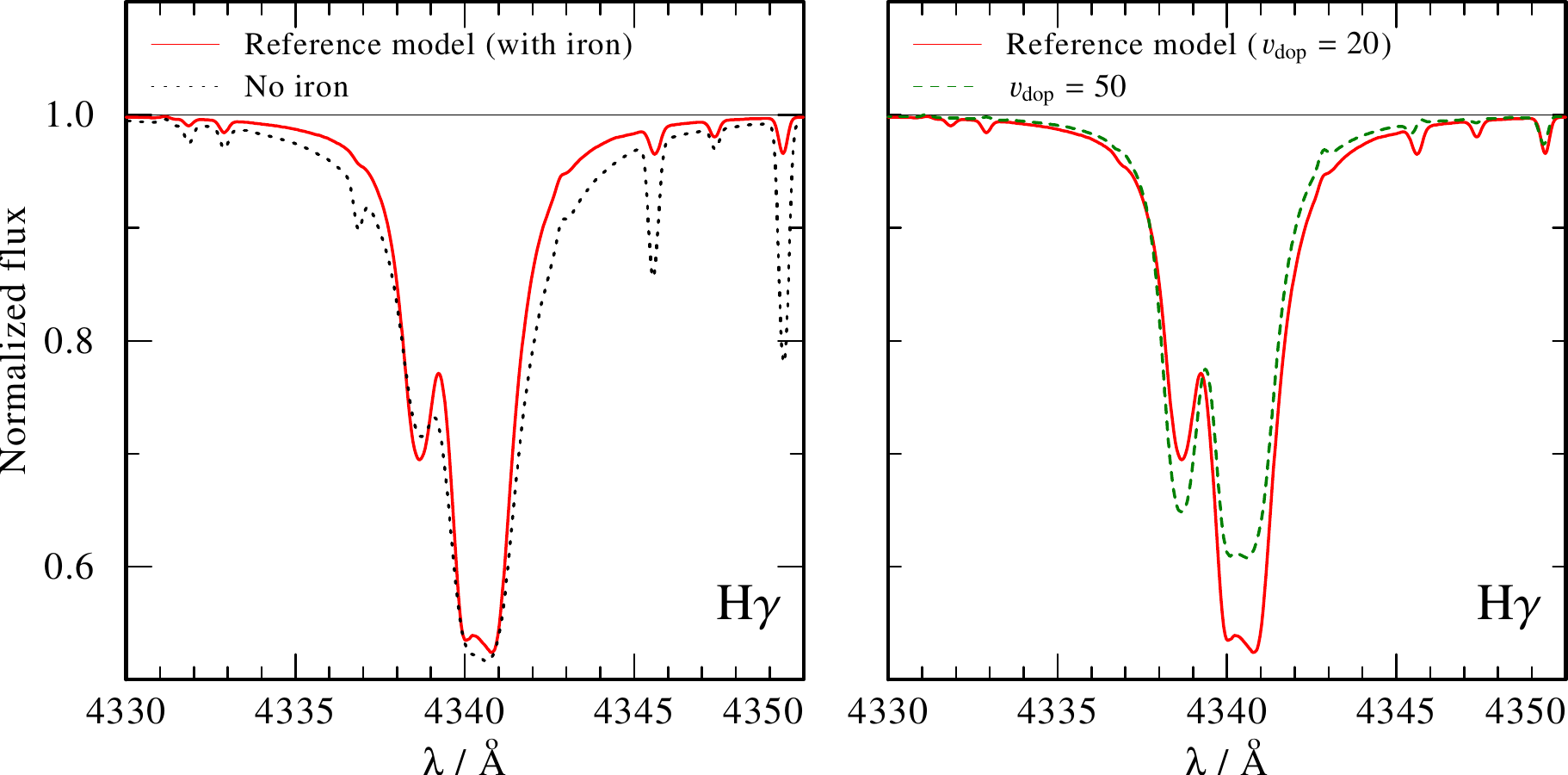}}
  \caption{Left panel: the spectrum around H$\gamma$ for the hydrostatically
           consistent supergiant test model (red solid line) is compared 
           to a similar PoWR model calculated without iron group elements (black dotted line).
           Right panel: the same test model is calculated with a higher
           Doppler broadening velocity $\varv_\text{dop}$ (green dashed line)
           and compared to the original.}
  \label{fig:vdopironcmp}
\end{figure}

        \subsection{Blanketing and Doppler velocity influence}
        \label{sec:resotherparam}

  It is unfortunate that the spectroscopic mass greatly depends on various parameters adopted 
  in the calculation. Two important examples will be discussed here. The first is the importance
  of the iron group elements. Since they are a dominant source for opacity in the atmosphere 
  of a massive star, their abundances significantly affect the radiative pressure, and thus the 
  inferred spectroscopic mass. Therefore iron group elements have to be included in model
  atmosphere calculations, even if these models are only used to analyse spectral regimes 
  without visible iron lines. The effect, which is called ``line blanketing'', does not
  only affect the temperature stratification, but also the density structure of a stellar atmosphere.
  This is illustrated in left panel of Fig.\,\ref{fig:vdopironcmp}, where we compare
  the H$\gamma$ line from our supergiant model (see Table\,\ref{tab:testmodels} for parameters) 
  to that of an identical model without iron. It impressively 
  demonstrates that neglecting elements, which greatly contribute to the total opacity, is not legitimate.
  When the iron elements are not included, the radiative pressure is smaller, and the line wings appear
  broader in the synthetic spectrum, leading to an underestimation of the gravity by $\approx 0.3$\,dex.
  These elements should thus be included in any consistent calculation, even if there is no
  observational abundance indicator. Using ``typical'' abundances from the local region or similar
  objects will usually cause a smaller error than neglecting such elements completely.

  Another quantity that is of major influence on the spectral appearance of OB star atmosphere
  models is the adopted Doppler broadening velocity $\varv_\text{dop}$. This velocity is
  used in the comoving frame calculations and reflects the combined influence of the thermal
  and microturbulent velocities. While the thermal velocity is calculated
  for each element as a depth-dependent manner in the formal integral, the comoving frame calculations are currently using a 
  constant $\varv_\text{dop}$
  for all elements. For the spectral appearance of Wolf-Rayet atmospheres the value of $\varv_\text{dop}$ 
  is of minor importance, but it has a notable impact on the spectra of our OB models, as it 
  affects the total radiative pressure in the atmosphere. This is illustrated in the 
  right panel of Fig.\,\ref{fig:vdopironcmp}, where we again compare the region around H$\gamma$ 
  for our supergiant test model, but now compare with a model that has been calculated with a larger Doppler velocity. 
  In the O-star regime, the value of $\varv_\text{dop} = 20\,$km/s has proved 
  to be sufficient. This demonstrates that it is imperative to choose a $\varv_\text{dop}$ reflecting
  the true thermal and turbulent velocity in the stellar atmosphere, despite the significantly longer 
  computing times in the comoving frame calculations that come along with smaller values of $\varv_\text{dop}$.

  \section{Comparison with TLUSTY}
    \label{sec:tlusty}
  
  The TLUSTY code \citep{Hubeny+1995} provides plane-parallel
  model atmospheres. The synthetic spectra that can be calculated from these models,
  e.g., with the SYNSPEC program from the same authors, are widely used for the 
  analysis of photospheric spectra of OB-type stars. In the following section, we   compare PoWR and TLUSTY model results. In all cases where spectra based on TLUSTY models
  are shown, they were obtained with the SYNSPEC code and are labeled as TLUSTY. 
  \footnote{The models from the TLUSTY O- and B-star grids (including
  their SYNSPEC spectra) can be obtained from the TLUSTY website
  at \texttt{http://nova.astro.umd.edu/}}
      
  While both PoWR and 
  TLUSTY are non-LTE codes, a major difference is that the PoWR models
  include the stellar wind.
  PoWR therefore adopts a spherical geometry, while TLUSTY  
  assumes a plane-parallel geometry. As thoroughly discussed by \citet{LH2003}, 
  the assumption of static, plane-parallel atmospheres is a fairly solid approximation for
  the photospheres of OB stars, which show negligible 
  signs of stellar winds and curvature effects. One can therefore expect that 
  in the limit of negligible mass-loss rates and small scale heights (see Eq.\,\ref{eq:hconst}), 
  the PoWR spectra would be close to the TLUSTY results. 
  To check that PoWR models of OB-type stars with 
  negligible winds show a good agreement with the corresponding TLUSTY models, 
  we dedicate this section to a comparison of dwarf, giant, and supergiant PoWR 
  models with their TLUSTY counterparts, focusing on the pressure-broadened Balmer
  lines. In all cases, PoWR models with the
  consistently treated quasi-hydrostatic domain are used, i.e., those models that we referred to as
  type (b) above.
    
\begin{figure}[ht]
  \resizebox{\hsize}{!}{\includegraphics[angle=0]{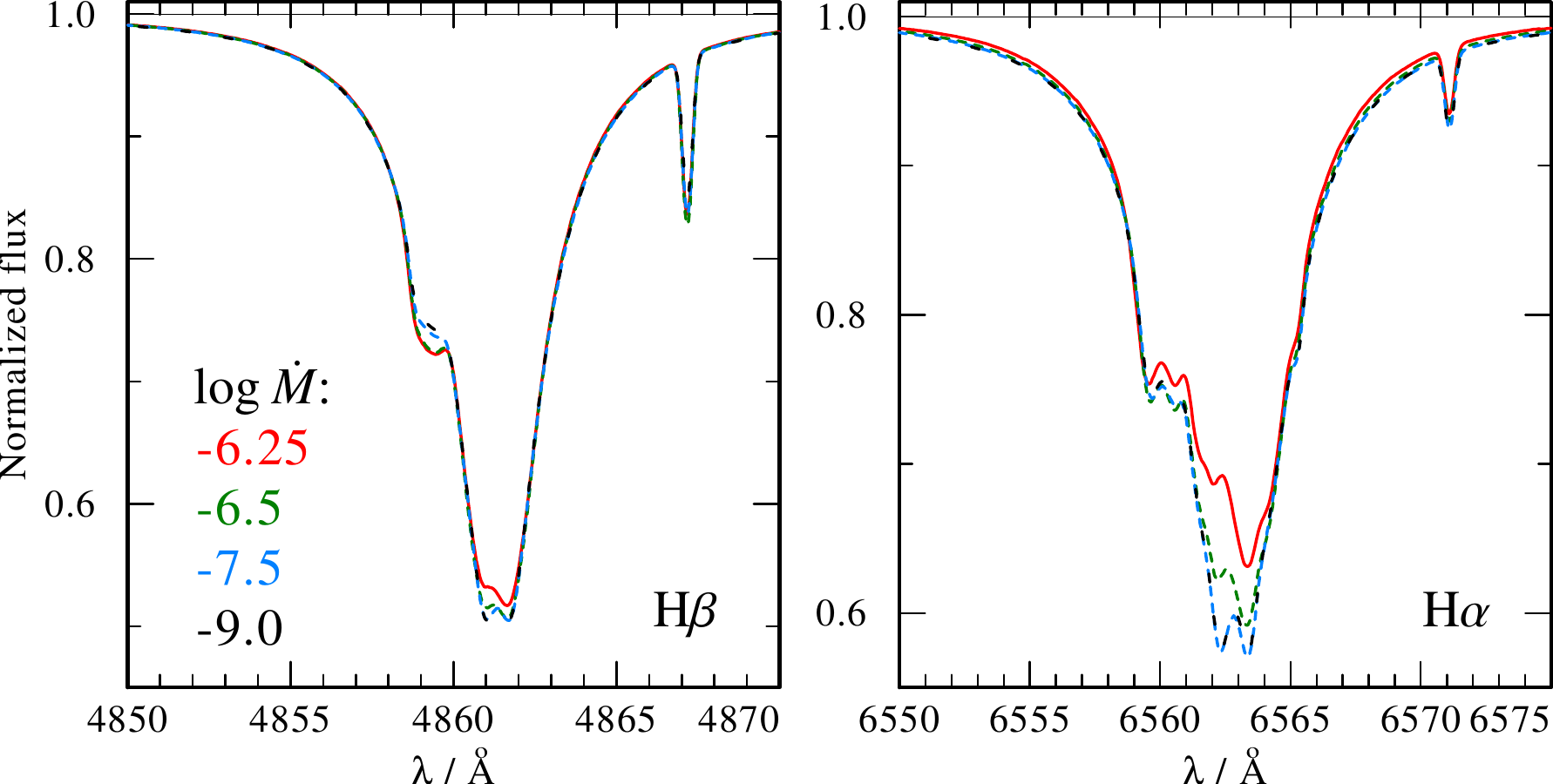}}
  \caption{Profiles of the H$\alpha$ and H$\beta$ lines from PoWR models with
           different $\dot{M}$. In the limit of low mass-loss rates, 
           the wind effect on the spectrum becomes
           negligible. For an easier comparison of the line wings, the electron 
           redistribution (see Fig.\,\ref{fig:redis}) is switched off in
           these simulations.}
  \label{fig:mdotinfluence}
\end{figure}

  Since we calcuated the models used in Sect.\,\ref{sec:results}  with ``typical'' O-star mass-loss
  rates, they are not expected to provide a very close agreement with the TLUSTY models. The
  discrepancies are most prominent in the H$\alpha$ line 
  for the supergiant and giant models, while small differences are also found in the 
  other Balmer lines. In Fig.\,\ref{fig:mdotinfluence}, we illustrate the effect of reducing 
  the mass-loss rate. It is evident that the spectral appearance converges in the limit of 
  small mass loss, and it is at this limit where we expect to obtain the closest agreement 
  with TLUSTY models. Furthermore, the sequence of models illustrates 
  that the absence of emission lines is not
  sufficient to deduce that mass loss is negligible.
  
\begin{figure}[ht]
  \resizebox{\hsize}{!}{\includegraphics[angle=0]{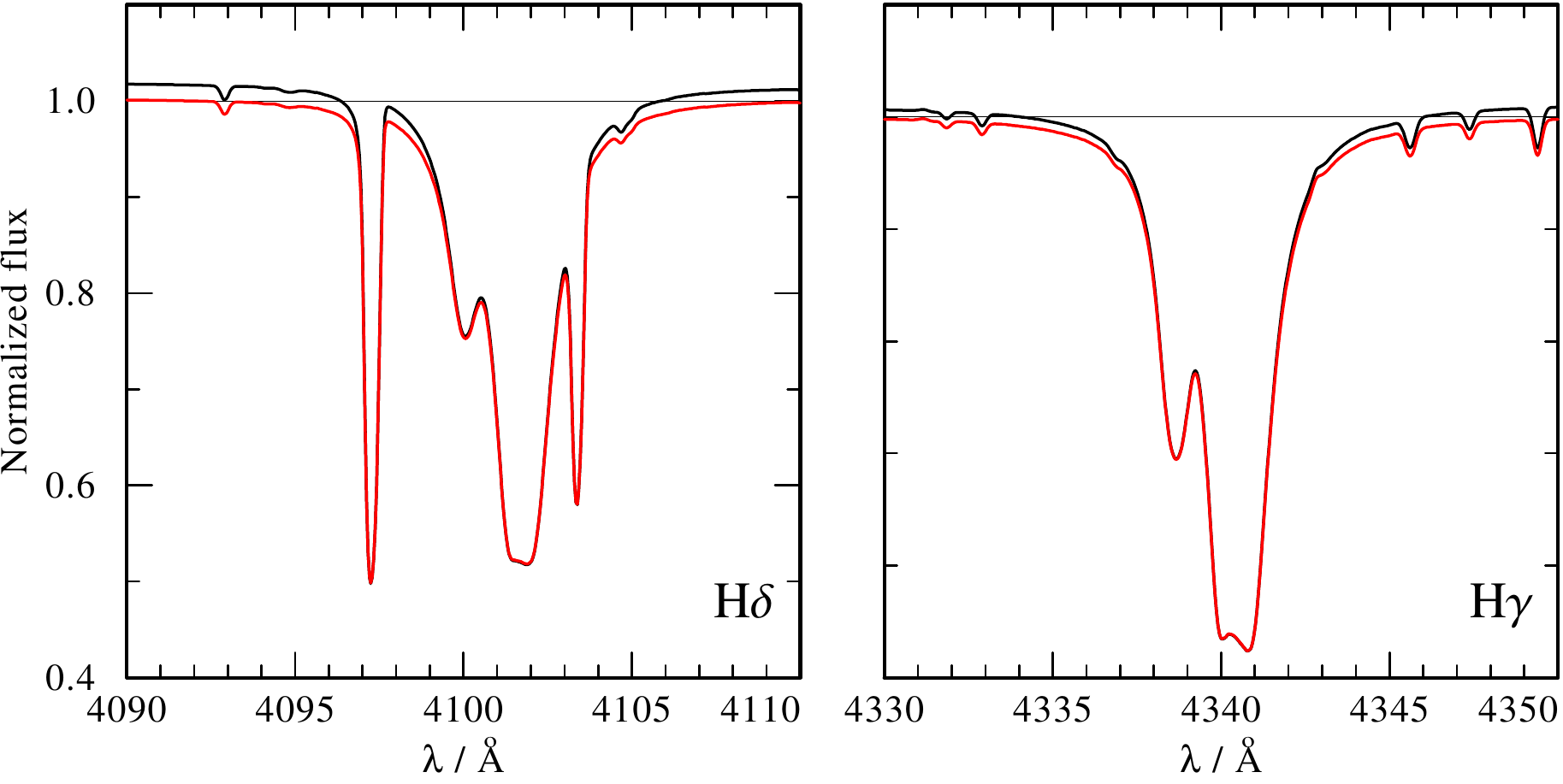}}
  \caption{Line profiles of H$\gamma$ and H$\delta$ for the same supergiant
           PoWR model. The red line denotes the spectrum that is obtained
           when switching off the electron redistribution in the formal integral,
           while the black line indicates the ``normal'' output, which includes
           the redistribution of free electrons.}
  \label{fig:redis}
\end{figure}

  Another PoWR feature, which would hinder a comparison with the TLUSTY models, 
  is frequency redistribution by Thomson scattering \citep{Mihalas1978Book}. Given 
  their large thermal velocities (${\sim}1000\,$km\,${\rm s}^{-1}$), free electrons can scatter 
  photons to significantly different wavelengths. Hence, photons that are trapped in an
  optically thick line core can be Doppler-shifted to the line wing or adjacent continuum,
  from which they can freely emerge and become visible as an excess emission.
  Especially at lower surface gravities and in spectral domains with a high density of spectral lines,
  the effect of frequency redistribution often results in a noticeable pseudocontinuum. When 
  the synthetic flux is normalized relative to the continuum, this has the appearance 
  of a continuum offset in the normalized spectrum. To illustrate this 
  effect, Fig.\,\ref{fig:redis} shows the H$\delta$ and H$\gamma$ lines (left and right 
  panels, respectively) for the consistent supergiant model
  including frequency redistribution (black solid line) and without (red solid line).
  Since TLUSTY 
  does not account for this effect, we disable it
  for the sake of comparison. However, we stress that the importance of this effect 
  has been demonstrated in various studies\footnote{It may seem arbitrary to 
  rectify the synthetic flux using the continuum before accounting for the redistribution. 
  However, the redistributed flux varies significantly around spectral lines, while 
  the ``unredistributed continuum'' is a slowly varying function of 
  $\lambda$ and therefore much more appropriate for normalization. Moreover, electron
  scattering contains vital information regarding 
  the physics in the stellar atmosphere, and should not be removed by normalization.}
  \citep{Hummer+1967, Auer+1968}. 

\begin{figure*}[ht]
  \resizebox{\hsize}{!}{\includegraphics[angle=-90]{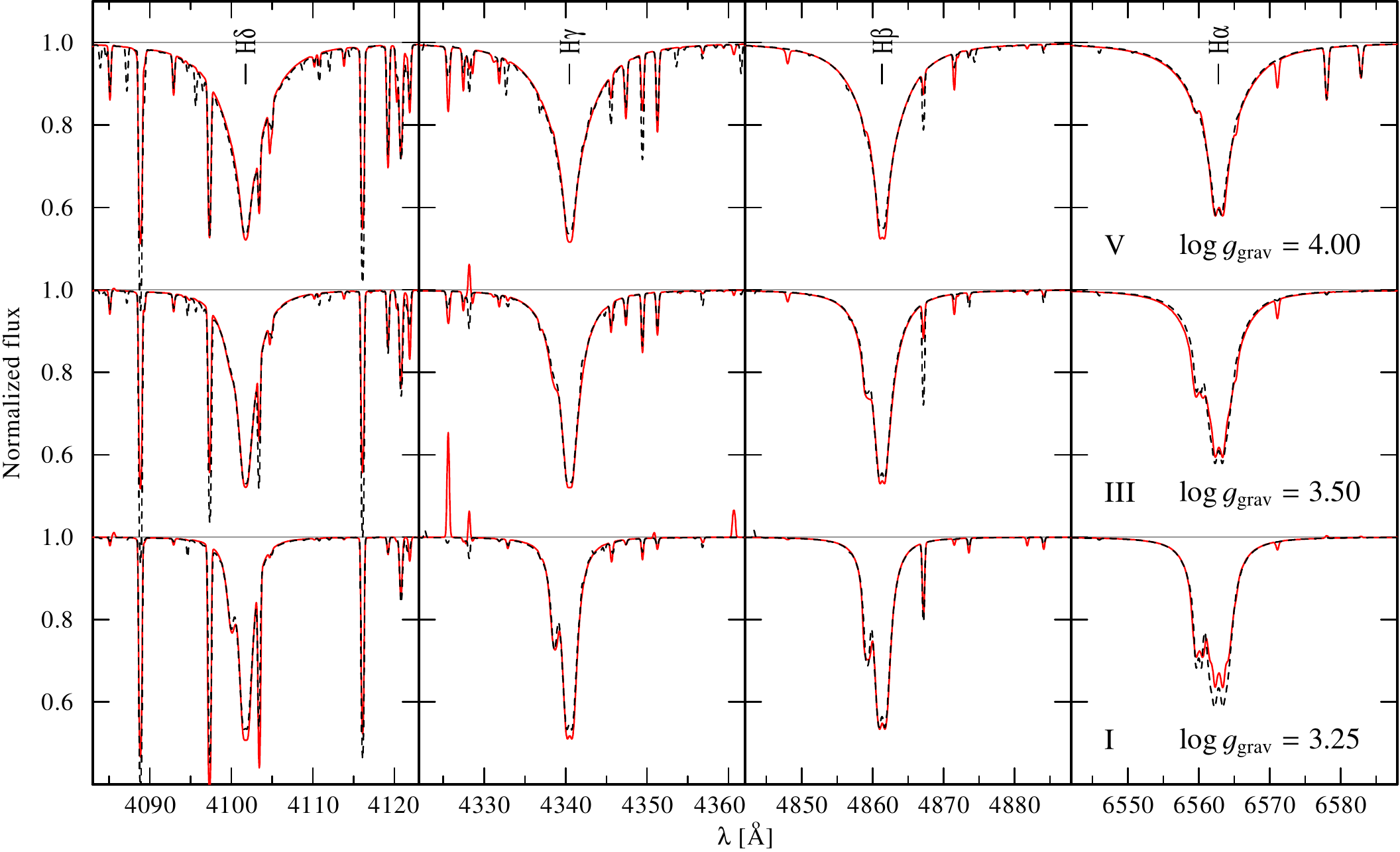}}
  \caption{Balmer line comparison between the PoWR models 
           (red, solid line) and the TLUSTY models (black, dashed line)
           for dwarfs (upper panels), giants (middle panels), and supergiants (lower panels). The 
           stellar parameters used in the PoWR models are as in Table\,\ref{tab:testmodels}, but with negligible 
           mass-loss rates ($\log \dot{M} = -9\,[M_\odot\,{\rm yr}^{-1}]$), small terminal velocities (500\,km\,${\rm s}^{-1}$), 
           and large radii, to avoid wind and curvature effects.}
  \label{fig:tlustycomp}
\end{figure*}

  The upper, middle, and bottom panels of Fig.\,\ref{fig:tlustycomp} show a comparison between the TLUSTY 
  (black, dashed line) and PoWR (red, solid line) for the
  dwarf, giant, and supergiant models, respectively,
  where we compare the first four Balmer members (from left to right: H$\delta$, 
  H$\gamma$, H$\beta$, H$\alpha$). The PoWR model parameters are identical to 
  those compiled in Table\,\ref{tab:testmodels}, but with a negligible mass-loss rate 
  of $\log \dot{M} = -9.0$\,$[M_\odot\,{\rm yr}^{-1}]$ and a small terminal velocity of $500\,$km\,${\rm s}^{-1}$. Furthermore,
  the radii of all stars were set to large values to diminish possible curvature effects and to approach a plane-parallel
  geometry.

  It is evident that a very good agreement in the line wings is obtained between the PoWR and TLUSTY models 
  for all three luminosity classes. In particular, the wings of the Balmer lines are 
  barely distinguishable. Hence, the application of either models would result in 
  practically the same surface gravity, provided that the mass loss is really negligible. 
  Interestingly, pressure broadening in the
  TLUSTY O-grid models from \citet{LH2003} uses analytical approximations to numerical 
  calculations \citep[cf.][appendix B]{Hubeny+1994}, 
  while PoWR performs interpolation over pressure broadening tables \citep{Lemke1997} for hydrogen lines.
  Given the good agreement, both methods seem to be adequate for the studied parameter regime,
  i.e., nondegenerated stars.

  
  While a comprehensive comparison between PoWR and TLUSTY along all spectral lines 
  is beyond the scope of the current paper, we note that there are significant
  differences regarding helium and metal lines. Not only does
  their strength differ by up to a factor of two between TLUSTY and PoWR, but 
  sometimes -- as visible in the supergiant comparison --
  they may even appear in emission from one code, and absorption from the other. Because of the
  non-LTE conditions, even small differences in the stratification can have large effects on such 
  weak lines, e.g.,\ between 
  two PoWR models differing only slightly in their temperature or gravity. 
  Indeed, we find similar discrepancies in the small lines. This can be seen in 
  Fig.\,\ref{fig:gravfix-multi}, where we compare the $\overline{\Gamma}_\text{rad}$- and 
  $\Gamma_\text{e}$-model spectra. Therefore one has to be careful deducing parameters,
  e.g., abundances, from only one particular line. Instead, several lines of
  more than one ionization stage should be compared if available.  

  
  
  
  
  \section{Summary and conclusions}
    \label{sec:conclusions}
 
    We present a set of stellar atmosphere models calculated with
    the most recent version of the PoWR code, using different approaches for the quasi-hydrostatic regime.
    In the limit of small mass-loss rates, we also compared the PoWR models 
    to the plane-parallel TLUSTY atmospheres.

    We conclude that a proper treatment of the quasi-hydrostatic regime is imperative for OB-type star
    modeling. For a consistent solution, the full radiative acceleration has to be 
    taken into account, including line and continuum contributions. 
    
    The spectroscopic masses will be 
    severely underestimated, by a factor of roughly 2, if models are
    used that account only for the radiation pressure on free electrons in
    the quasi-hydrostatic domain. This holds for all luminosity classes.
    
    Omitting elements that significantly contribute to the total opacity
          compromises the density stratification in the quasi-hydrostatic domain, and 
          consequently leads to an
          inconsistent stellar masses. Specifically neglecting important elements, such as Fe, 
          in the models is by no means legitimate.
    
    The effect of mass loss on the spectra of typical OB stars may seem subtle, 
          but even small mass-loss rates can change the line profile and thus
          might affect deduced gravities. The absence of emission lines in an
          observation does not imply negligible mass loss.
    
    In the limit of small mass-loss rates and vanishing curvature effects, the 
          emergent spectra of the PoWR model 
          atmospheres generally agree very 
          well with TLUSTY models calculated with the same stellar parameters.

\begin{acknowledgements}
  We would like to thank the anonymous referee for the fruitful suggestions that
  helped to significantly improve this paper. We would also like to thank Ivan Hubeny for the 
  interesting and productive discussions that lead to the comparisons presented 
  in this work. The first author of this work (A.S.) is supported by the Deutsche 
  Forschungsgemeinschaft (DFG) under grant HA 1455/22. T.S. is grateful
  for financial support from the Leibniz Graduate School for Quantitative
  Spectroscopy in Astrophysics, a joint project of the Leibniz Institute
  for Astrophysics Potsdam (AIP) and the Institute of Physics and Astronomy
  of the University of Potsdam. A.S. would like to thank the 
  Aspen Center for Physics and the NSF Grant \#1066293 for hospitality 
  during the invention and writing of this paper.
\end{acknowledgements}


\bibliographystyle{aa} 
\bibliography{hydrostatpaper}





\end{document}